\documentclass{elsart}
\usepackage{amsmath}
\usepackage{amssymb}
\usepackage{graphics}

\begin{document}

\begin{frontmatter}
\title{Avalanches in complex spin networks}

\author[agh-ust]{K. Malarz\corauthref{km1}},
\corauth[km1]{Corresponding author. Fax: +48 12 6340010}
\ead{malarz@agh.edu.pl}
\author[agh-ust]{W. Antosiewicz},
\author[agh-ust]{J. Karpi\'nska},
\author[agh-ust]{K. Ku{\l}akowski}
\and
\author[jsi]{B. Tadi\'c}
\address[agh-ust]{
Faculty of Physics and Applied Computer Science,
AGH University of Science and Technology,
al. Mickiewicza 30, PL-30059 Cracow, Poland
}
\address[jsi]{
Department for Theoretical Physics,
Jo\v{z}ef Stefan Institute,
P.O.Box 3000, SI-1001, Ljubljana, Slovenia
}

\begin{abstract}
We investigate the magnetization reversal processes on classes of 
complex spin networks with antiferromagnetic interaction along the network links. 
With slow field ramping the hysteresis loop and avalanches of spin flips occur 
due to topological inhomogeneity of the network, even without any disorder
of the magnetic interaction [B. Tadi\'c {\em et al.}, Phys. Rev. Lett. {\bf 94} (2005) 137204].
Here we study in detail properties of the magnetization 
avalanches, hysteresis curves and density of domain walls and show how they can 
be related to the structural inhomogeneity of the network. The probability 
distribution of the avalanche size, $N_s(s)$, displays the power-law behavior for small $s$,
i.e. $N_s(s)\propto s^{-\alpha}$. For the scale-free networks, grown with preferential 
attachment, $\alpha$ increases with the connectivity parameter $M$ from $1.38$ for $M=1$ 
(trees) to 1.52 for $M=25$. For the exponential networks, $\alpha$ is close to 1.0
in the whole range of $M$.
\end{abstract}

\end{frontmatter}

\section{\label{sec1}Introduction}

Growing networks is a vivid area of interdisciplinary sciences, with long list 
of applications \cite{book,book2,ab,drm,nwm,past}. The mainstream of literature is 
concentrated on the structure of the networks. However, from the point of view of 
some applications it is of interest to decorate nodes with additional degrees of freedom. 
In the simplest case, these variables are discrete, with two possible values: $\pm 1$.
Such an extension can be useful when discussing numerous examples, from neural and logical 
networks through sexual networks to quantum gravitation \cite{extension}.
Our topic here is a network of interacting magnetic moments (spins), with two possible magnetic 
orientations, up or down. The models of spin networks with ferromagnetic interaction
have been considered in several recent studies \cite{ferro,solomon}. 
The case of spin networks with antiferromagnetic 
interactions is essentially different because of the frustration 
effect along closed cycles on the graph \cite{stak,BT-etal}.
This difference is well known also for regular lattices \cite{galam}.

Recently we introduced and investigated field-driven dynamics of  spins on complex networks with 
the antiferromagnetic interaction \cite{BT-etal}. We found that the structural complexity of the 
networks leads to avalanches of spin flips and a criticality of the hysteresis loop.
The desired shape of hysteresis curves can be obtained by tuning the clustering parameter $M$.
This parameter counts the number of links, which attach each new node to the growing network.

In this work we expand the study of the reversal processes in the antiferromagnetic spin networks.
We demonstrate the genesis of avalanches due to inhomogeneity
in network's connectivity when the clustering is low. 
In this limit the avalanche structure and duration can be clearly interpreted in terms of the theoretical distribution of connectivity and depth of the network.
For large clustering, the character of the avalanche spectra $N_s(s)$ for large $s$ change from mulitipeaky to a soft decreasing.
For small avalanches, the avalanche spectra display the power-law character in the whole range of the clustering parameter $M$. 
We define the size $s$ of avalanche as a number of spins which were flipped at a given value of external field $H$.

In Section \ref{sec2} we introduce details of the model and discuss spin reversal on small graphs. 
Section \ref{sec3} contains details of the emergent structures for two classes of growing networks 
exponential and scale-free. Our results on the magnetization reversal in these networks are given 
in Section \ref{sec4} and discussed in Section \ref{sec5}.

\section{\label{sec2} Spin reversal on small graphs}

\begin{figure*}
\begin{center}
\begin{tabular}{llll}
a) \resizebox{0.35\hsize}{!}{\includegraphics{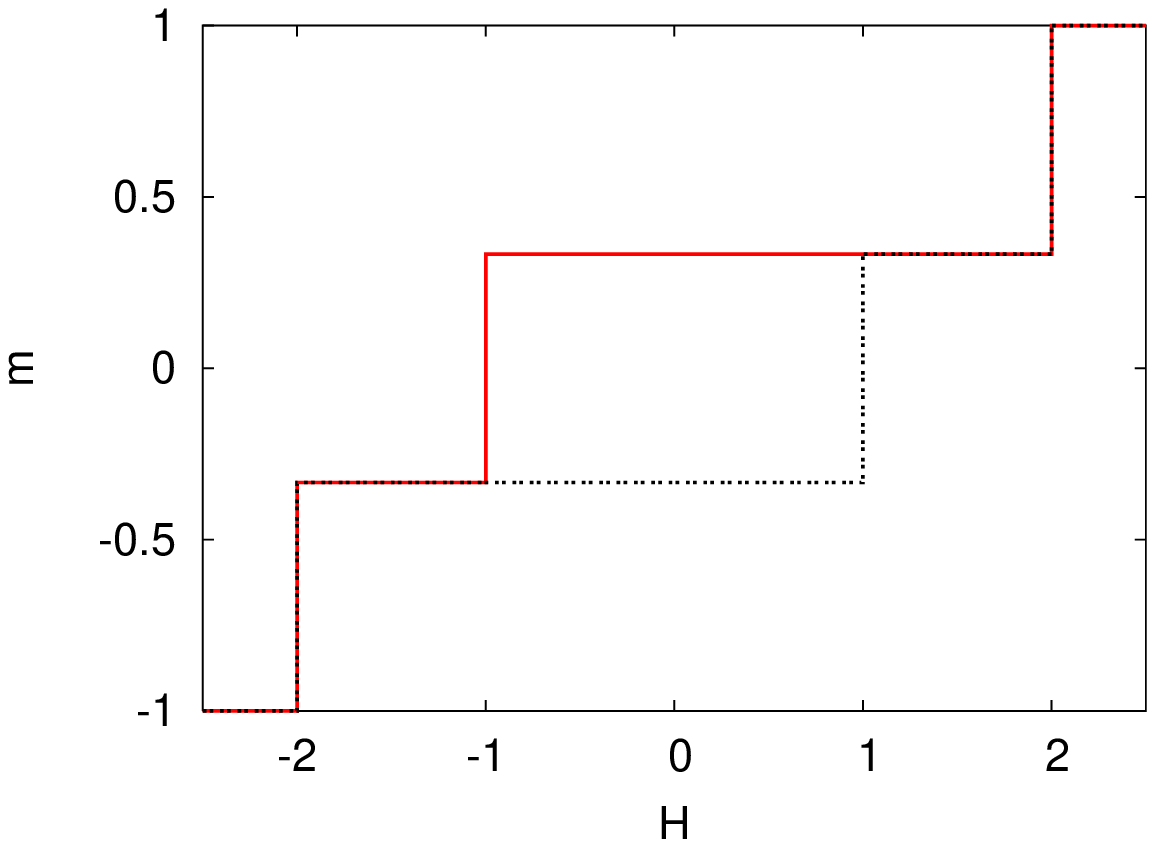}} &
\scalebox{0.12}{\includegraphics{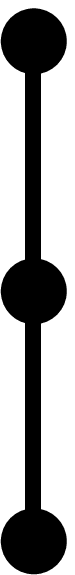}} &
b) \resizebox{0.35\hsize}{!}{\includegraphics{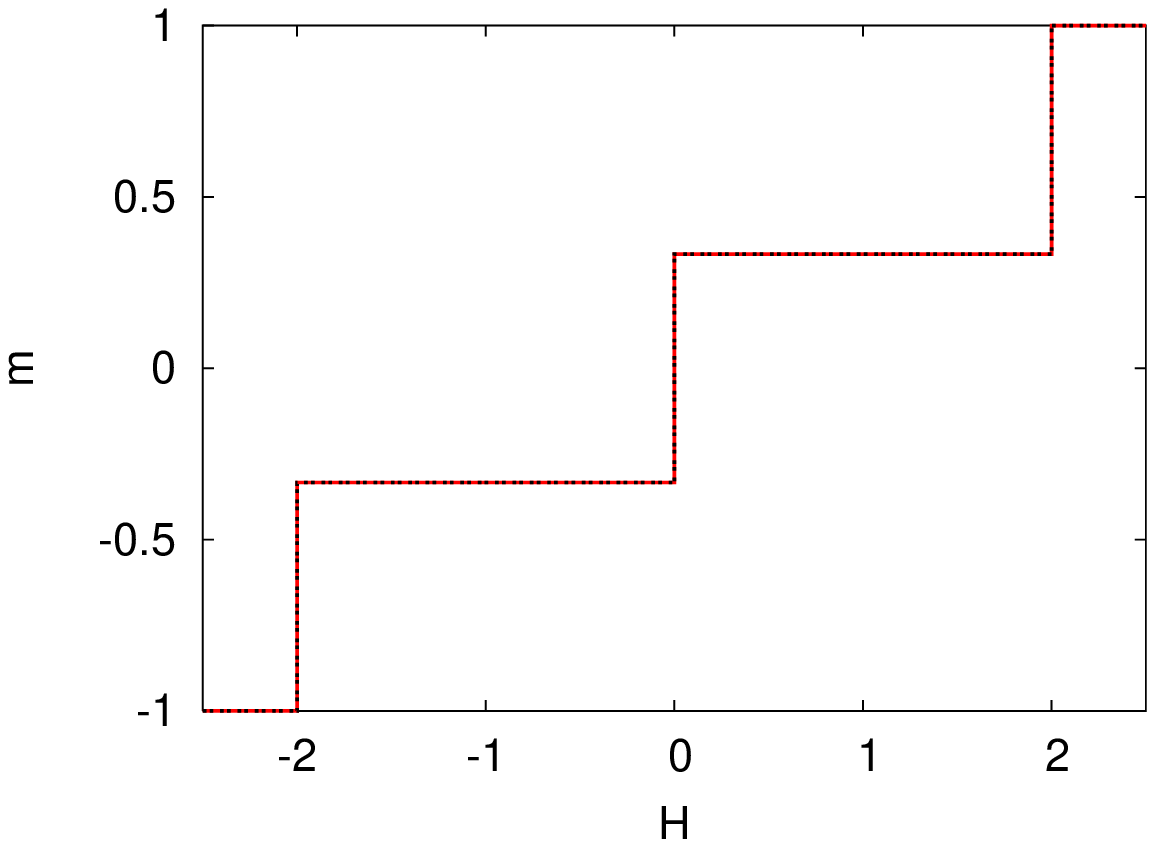}} &
\scalebox{0.12}{\includegraphics{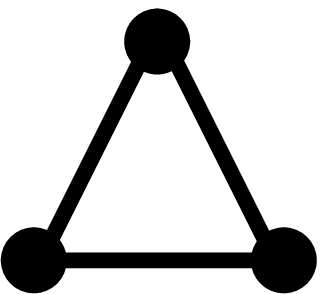}} \\
c) \resizebox{0.35\hsize}{!}{\includegraphics{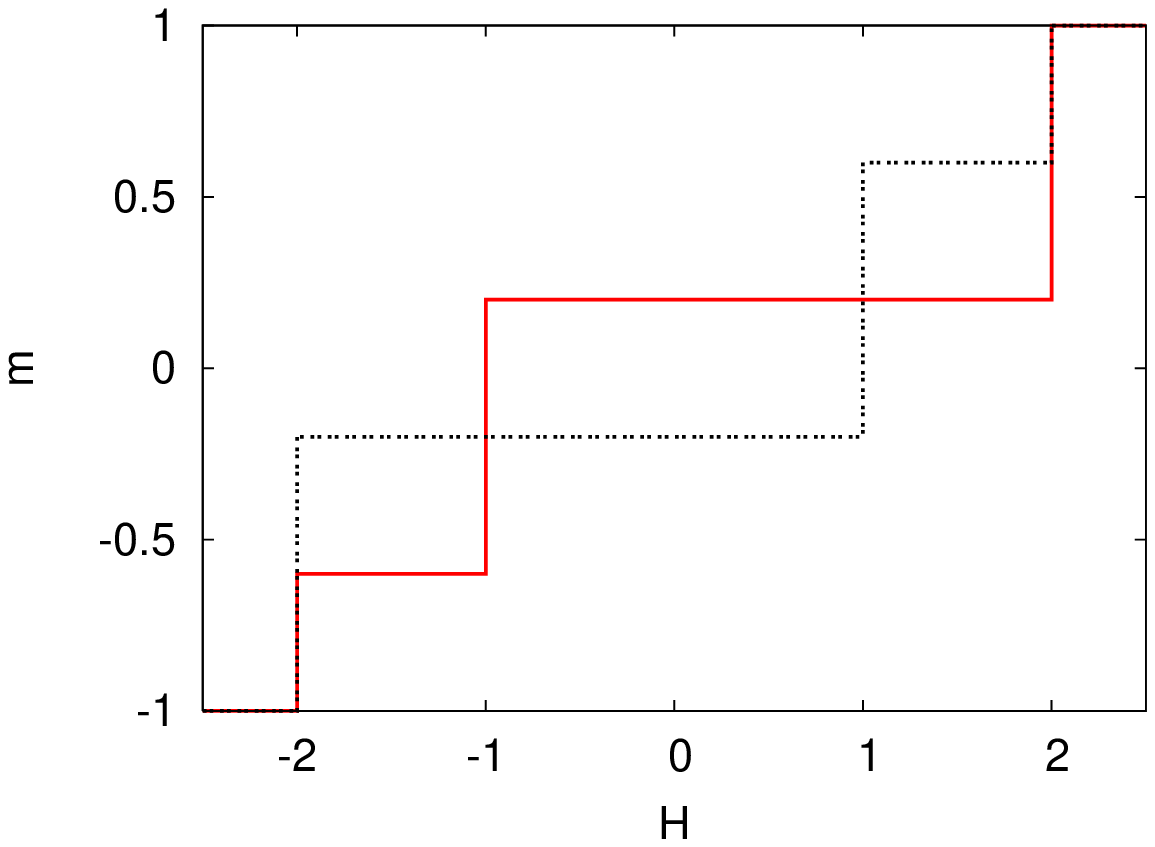}} &
\scalebox{0.12}{\includegraphics{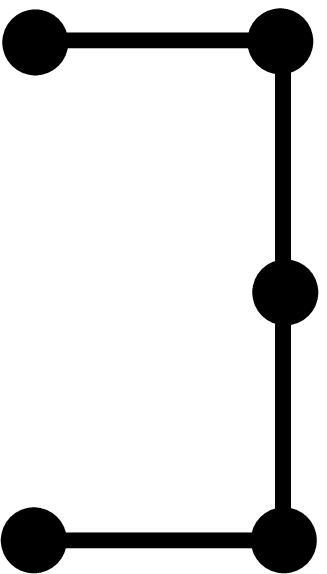}} &
d) \resizebox{0.35\hsize}{!}{\includegraphics{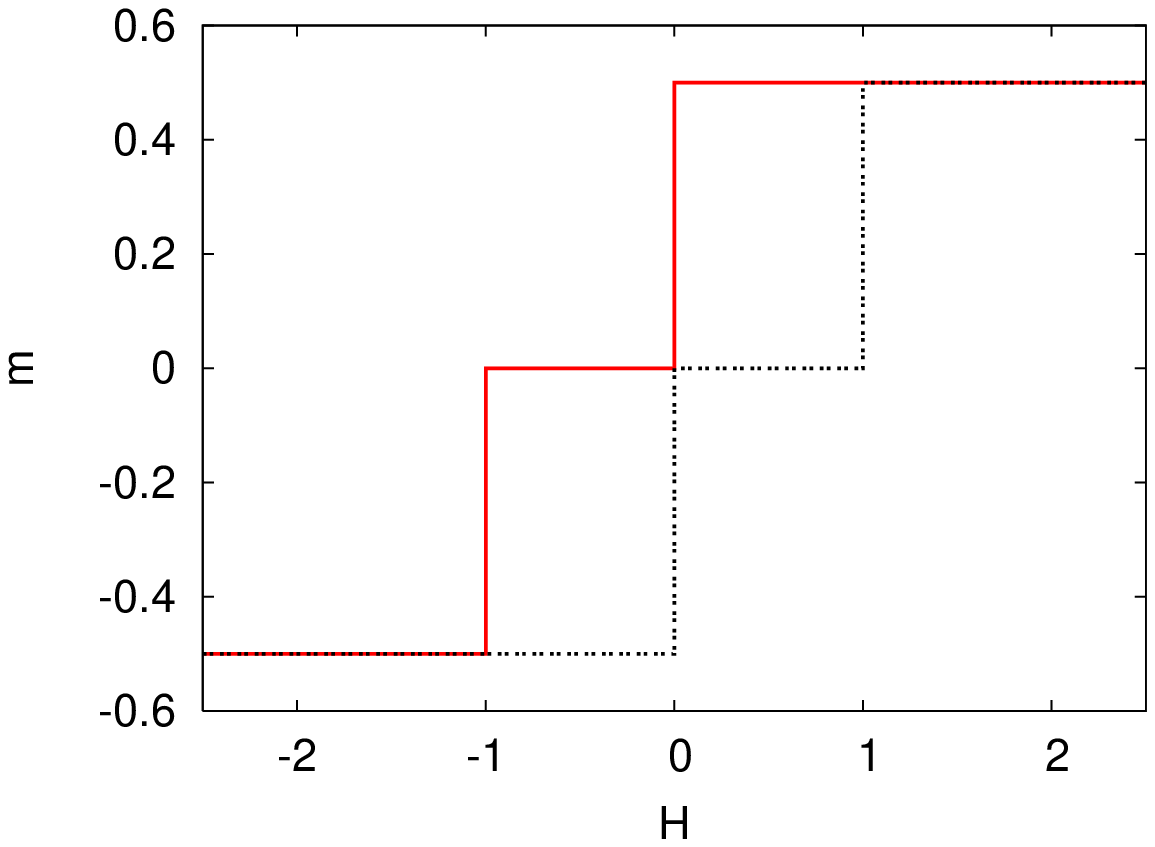}} &
\scalebox{0.12}{\includegraphics{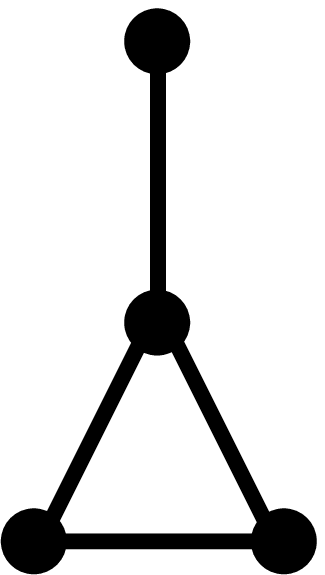}}\\
\end{tabular}
\end{center}
\caption{\label{fig-net-his} Toy graphs and their hysteresis loops for decreasing field (continuous red line) and increasing field (dotted black line).}
\end{figure*}

To describe our procedure in details, we begin with small simple graphs (Fig. \ref{fig-net-his}). A spin $S_i=\pm 1$
is placed at each node. If two nodes are linked together, antiferromagnetic interaction 
tends to keep them in one of two antiparallel configurations. Energy of the whole system is
written as
	\begin{equation}
	E= - J \sum_{\langle ij\rangle}S_i S_j - H\sum_iS_i,
	\end{equation}
where coupling constant $J=-1$ for the antiferromagnetic interaction and the first summation 
goes over all nearest neighbor pairs $\langle ij\rangle$.
We assume  symmetric neighbors, i.e. if spin $i$ is neighbor to spin $j$ then also $j$ is neighbor to $i$.
(We note, that results may be drastically different when the neighbor relations are directed \cite{solomon,directed}.)
To draw the hysteresis loops, graphs 
are placed in saturating magnetic field $H>0$, which forces all spins to be positive, i.e. to 
point upwards. 
We ramp the field $H$ by integer steps \cite{comment} from $n+\delta$ to $n-1+\delta$ 
in order to avoid ambiguous spin orientation at integer field values  $H=n$ ($n\in\mathbb{Z}$).
As the field decreases gradually at zero temperature $T=0$ some spins are flipped 
($S_i\to-S_i$) because of the 
antiferromagnetic interaction with their neighbors. 

In Fig. \ref{fig-net-his} we present some graphs and their hysteresis loops.
We note that for some of them, the obtained hysteresis loop depends on the order of updating.
As an example, consider a chain of five spins presented in Fig. \ref{fig-net-his}(c).
In saturation, effective fields at node 2, 3 and 4 are the same.
If the central spin 3 is updated at first, only this spin flips at field $H=-2J$.
If we update spins starting from the end of chain (as it is done here), the flipped ones 
are 2 and 4.
Then, not only the flipped spin labels but also the number of flipped spins is influenced
 by the order of updating.
However, any procedure suffers this ambiguity until we do not add a small variation of 
the magnetic moments or the interactions; in this case, the Pardavi-Horvath algorithm  
\cite{parda} is appropriate, where the field is changed continuously and a spin where 
the effective field is zero is flipped as first.
On the other hand, the synchronous scheme of spin updates may lead to persistent 
oscillations as  when $S_i\to-S_i$, and the avalanche will never stop.
Here we apply causal way of spin updating, i.e. from the spin decorating ``the oldest'' node of the graph towards the one assigned to the node added at very end.
Below we present results averaged over a large number $N_{\text{run}}$ of graphs 
with different shapes, and the influence of the order of updating is expected to be 
averaged out.

\begin{figure*}
\begin{center}
\resizebox{0.45\hsize}{!}{\includegraphics{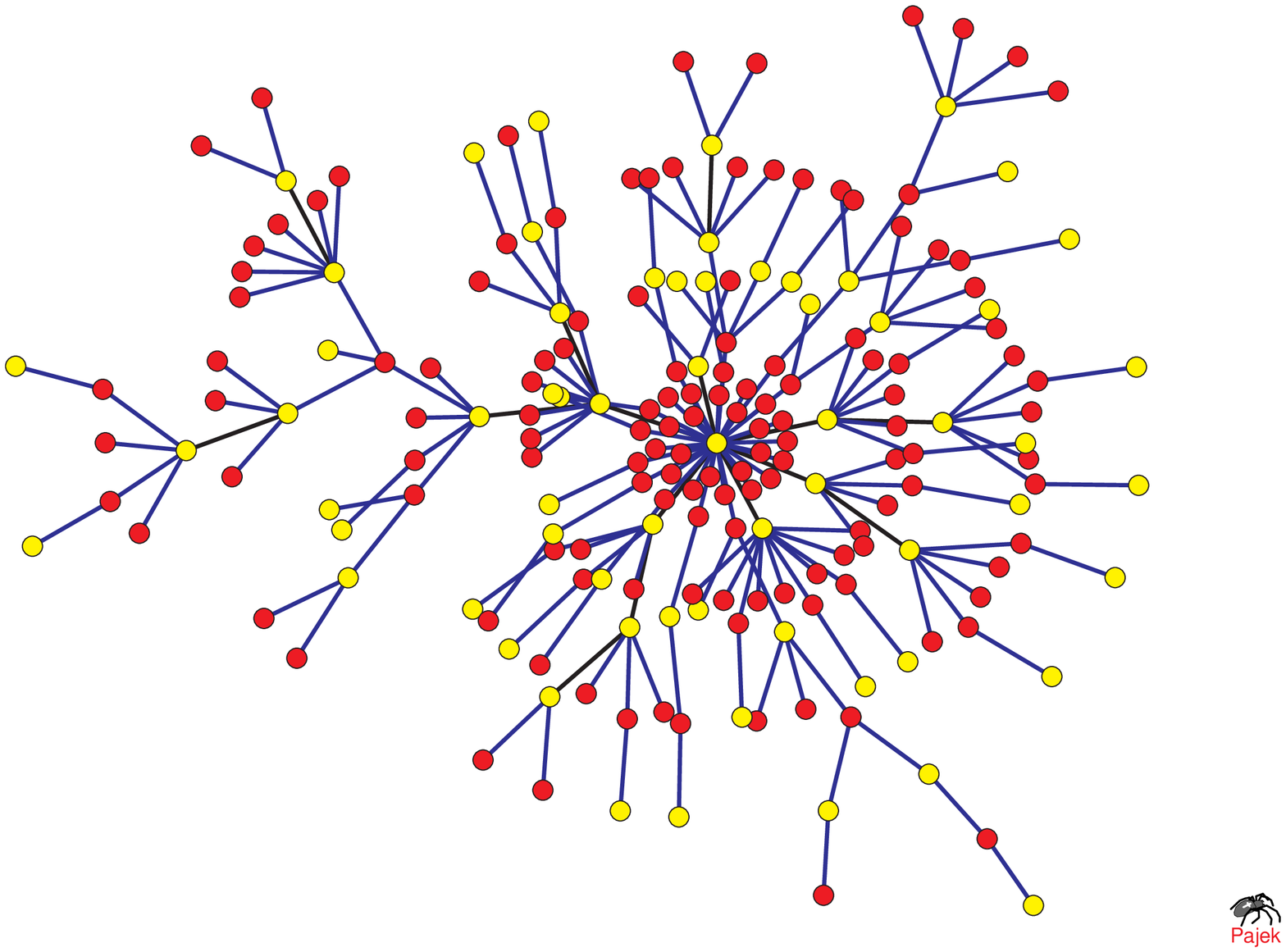}}
\resizebox{0.45\hsize}{!}{\includegraphics{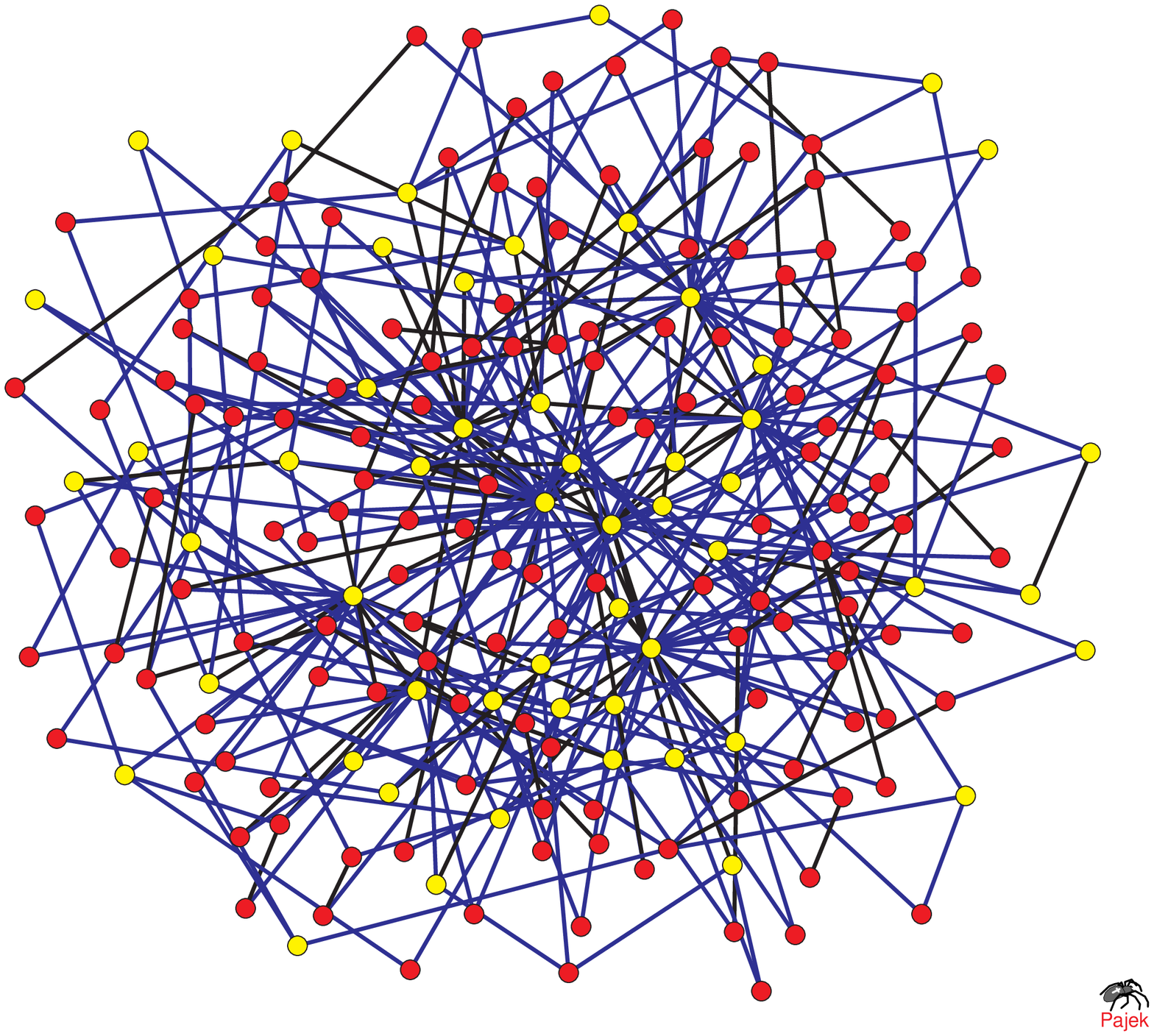}}
\end{center}
\caption{\label{fig-spin} (Color on-line). Snapshots from simulation for scale-free tree ($M=1$, left) and scale-free simple graph ($M=2$, right) for $N=200$ and magnetic field $H=0-\delta$.
Spins $S_i=-1$ (red) and $S_i=+1$ (yellow) and their pairs $S_iS_j=+1$ (black) $S_iS_j=-1$ (blue) for all $1\le i,j\le N$.
(Figures using Pajek \cite{pajek}.)}
\end{figure*}

\section{\label{sec3}Structure of networks}

It is well known that the structure of the growing network emerges in the process 
of node adding and linking.
Here we are interested in the growing exponential (EXN) and the scale-free (SFN) 
graphs \cite{drm}.
In both cases, a network grows by successive adding of new nodes.
Each new node is linked to $M$ different pre-existing nodes, which are selected randomly
 (EXN) or with preference (SFN).
Here we limit our interest to the case when the preferential probability of selecting a 
given node is proportional to 
the degree $k$ of this node, i.e. to the number of currently present  bonds at that node.
This procedure leads to the scale-free character of the degree distribution for SFN, i.e.  $P(k\ge M)\propto k^{-\gamma}$, where theoretically $\gamma=3.0$ for large network size $N$ independently on the parameter $M$ \cite{ab,drm,nwm,ba}.
For EXN trees ($M=1$), $P(k\ge M)= 2^{-k}$ \cite{drm}, and for EXN with $M=2$ we get $P(k\ge M)=3/4\cdot(3/2)^{-k}$ \cite{app}.
The parameter $M$ can be tuned to modify the sparseness of the networks, from $M=1$ (trees) to higher values, where cyclic paths appear.
If the number of nodes in such a cycle is odd, as in a triangle, a frustration of spins occurs: some antiferromagnetic bonds join spins of the same orientation, and the minima of energy become shallow.
The clustering coefficient $C$ \cite{ab} depends on the node age. 
The clustering profiles, as shown in \cite{BT-etal} for the case  $M=5$, exhibits a power-law tail both for EXN and SFN. 
The networks with low clustering ($M=1,2$) reveal effects of linking inhomogeneity.

We stress that all bonds are antiferromagnetic and there is no bond disorder in the traditional sense. 
However, the structure of the network cannot be treated as ordered in the sense of periodic lattices. 
In most of graphs each node has its unique position in the network \cite{comm}.
Therefore, a question arises, if the spin glass phase is possible in the antiferromagnetic spin-networks with clustering $M>1$ at least at $T=0$. 
In terms of the accepted spin-glass terminology, the presence of nodes with large connectivity may be regarded  either as a long-range interaction effect, or as an equivalence of a large system dimensionality.
However, such structure is not translationally invariant \cite{krawczyk}.
The network shows a nontrivial connectivity profile, which implies new features of the spin-glass order and requires refined theoretical approaches \cite{nest}.

In Fig. \ref{fig-spin} examples of the  networks considered in this work are presented.
For clarity of the plots the network size is kept moderate.
In the simulations we use networks with $N=10^4$ nodes. 

\section{\label{sec4}Hysteresis loops and avalanche distributions}

\begin{figure}
\begin{center}
 \resizebox{0.45\hsize}{!}{\includegraphics{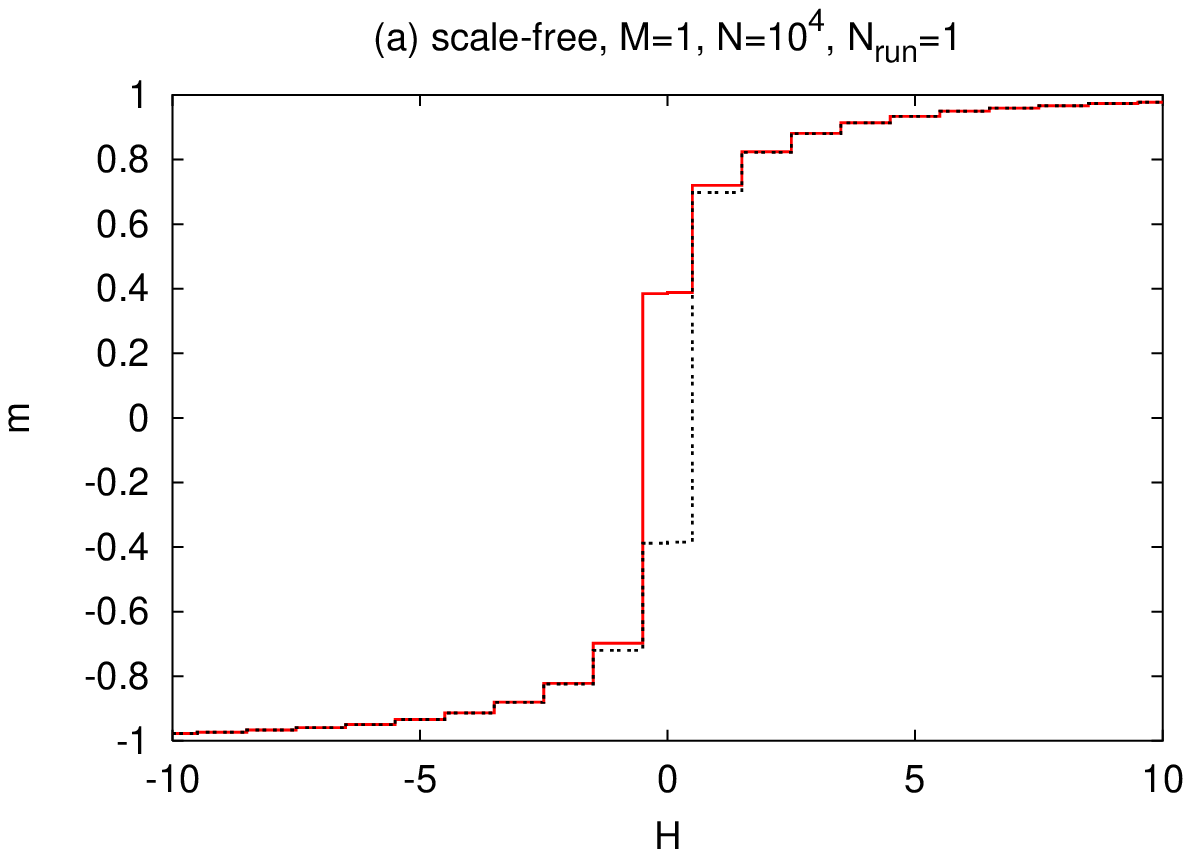}}
 \resizebox{0.45\hsize}{!}{\includegraphics{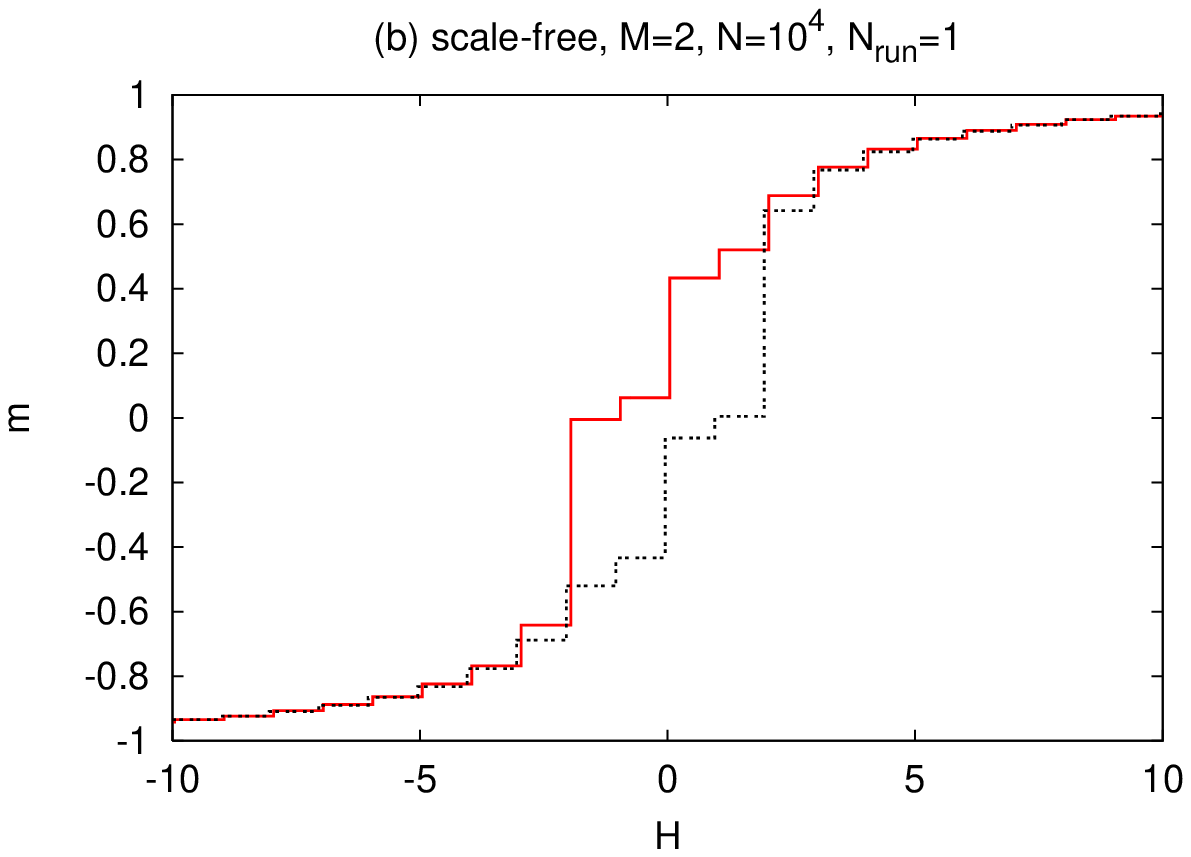}} \\
 \resizebox{0.45\hsize}{!}{\includegraphics{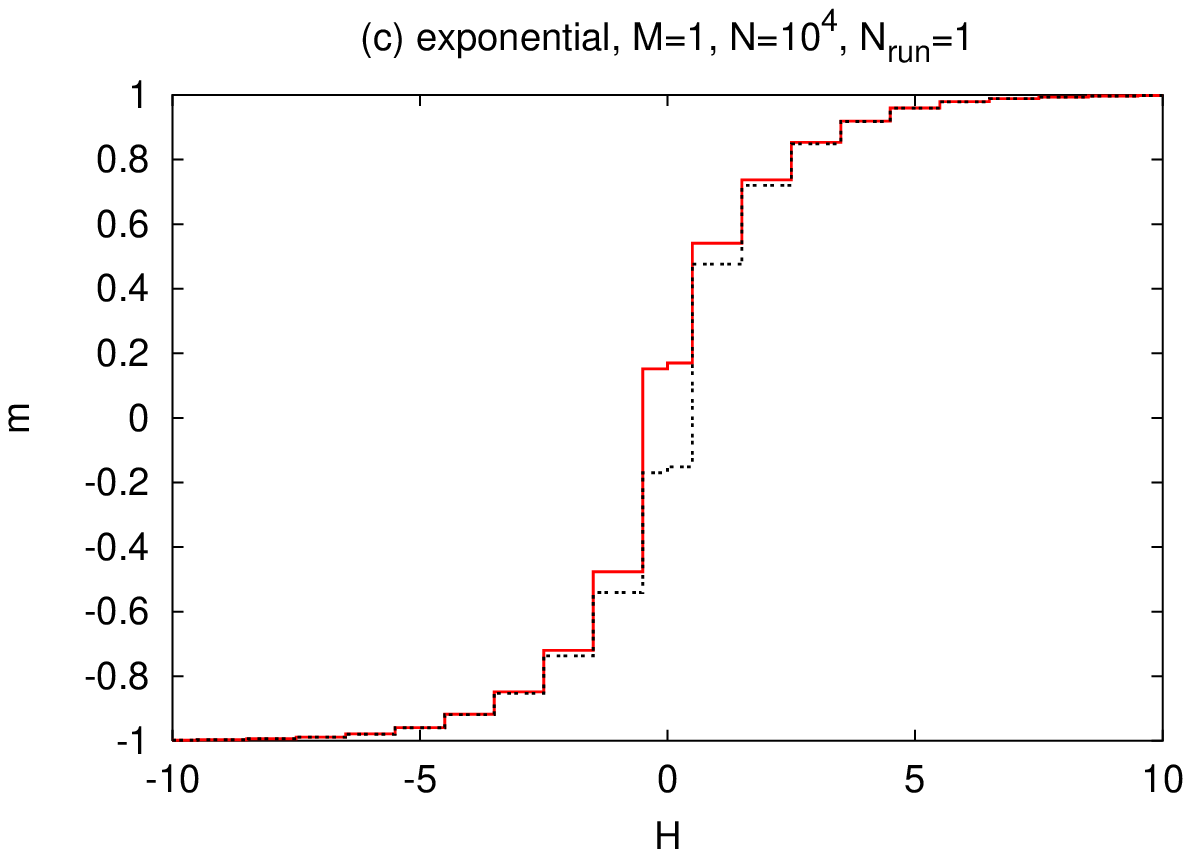}}
 \resizebox{0.45\hsize}{!}{\includegraphics{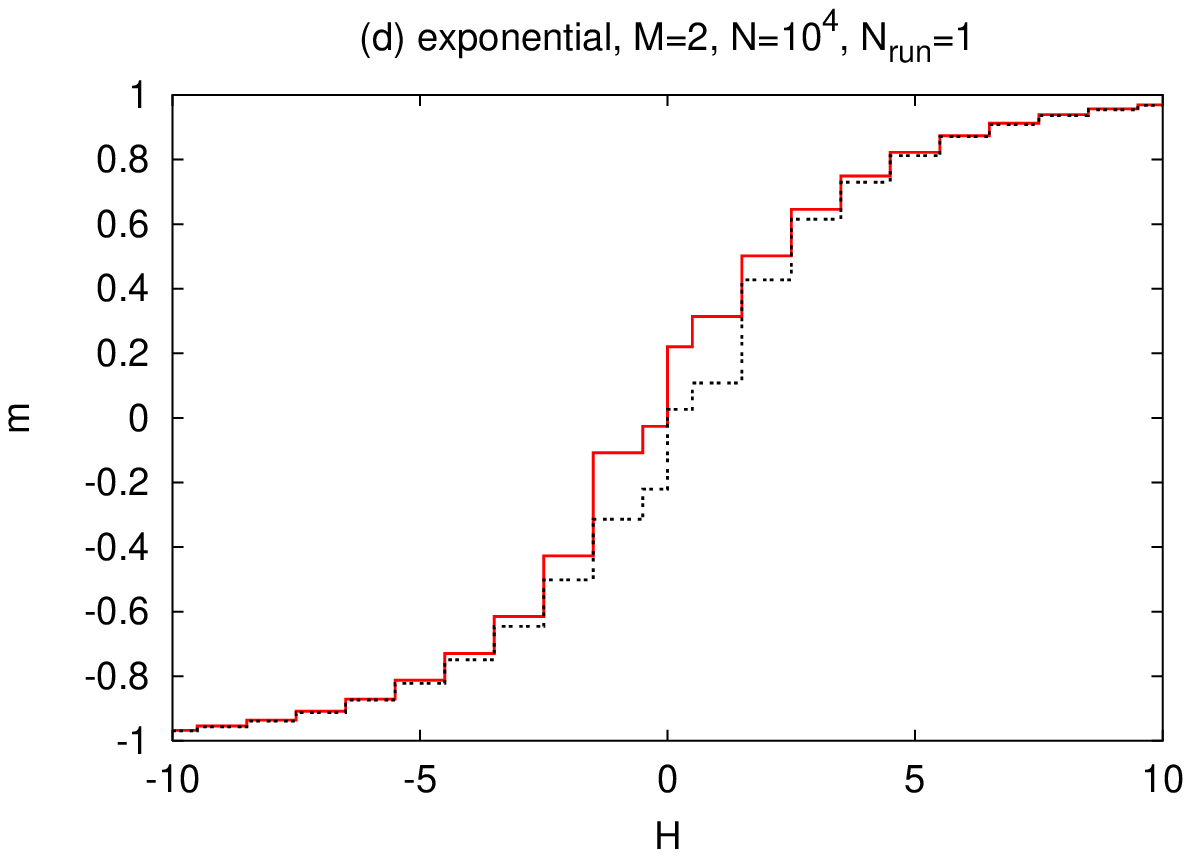}}
\end{center}
\caption{\label{fig-mH} The field dependence of the magnetization $m(H)$.
The field is expressed in $|J|$ units.}
\end{figure}

\begin{figure}
\begin{center}
\resizebox{0.80\hsize}{!}{\includegraphics{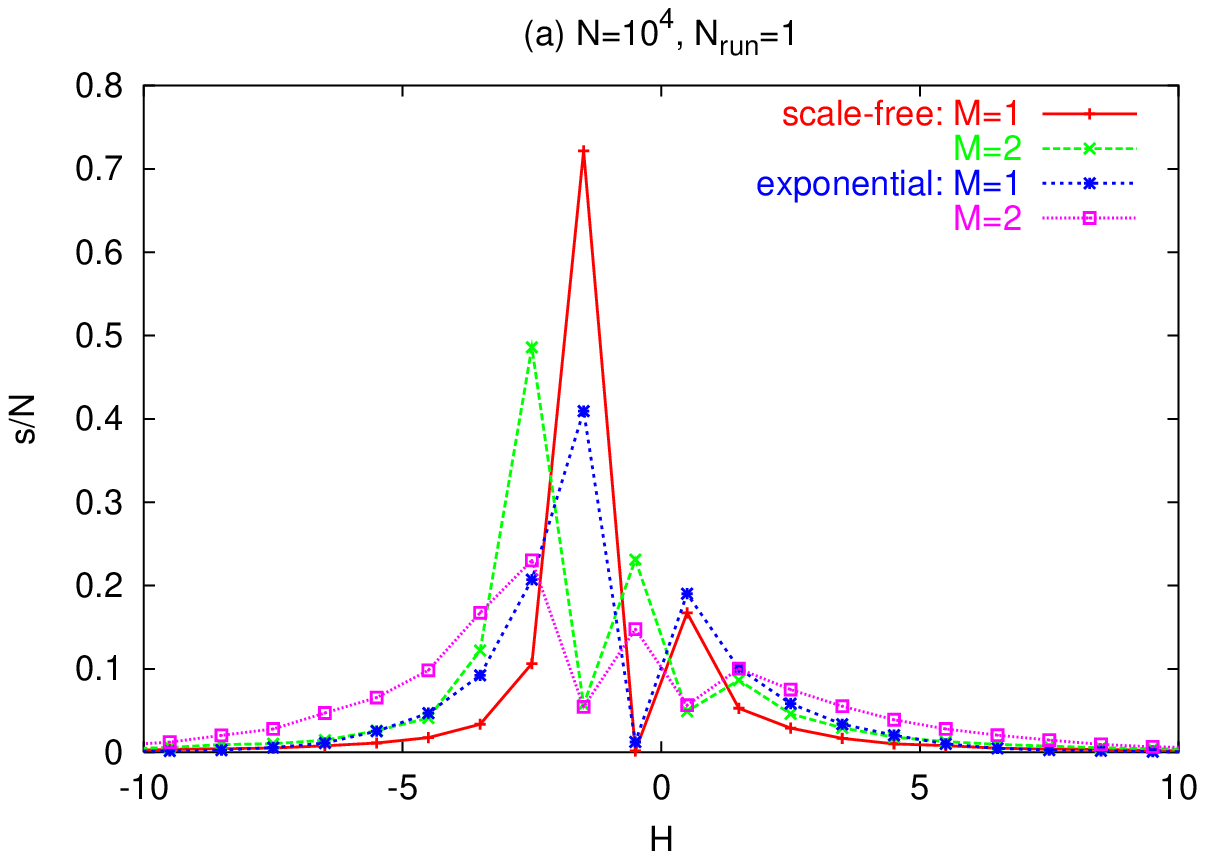}}\\
\resizebox{0.80\hsize}{!}{\includegraphics{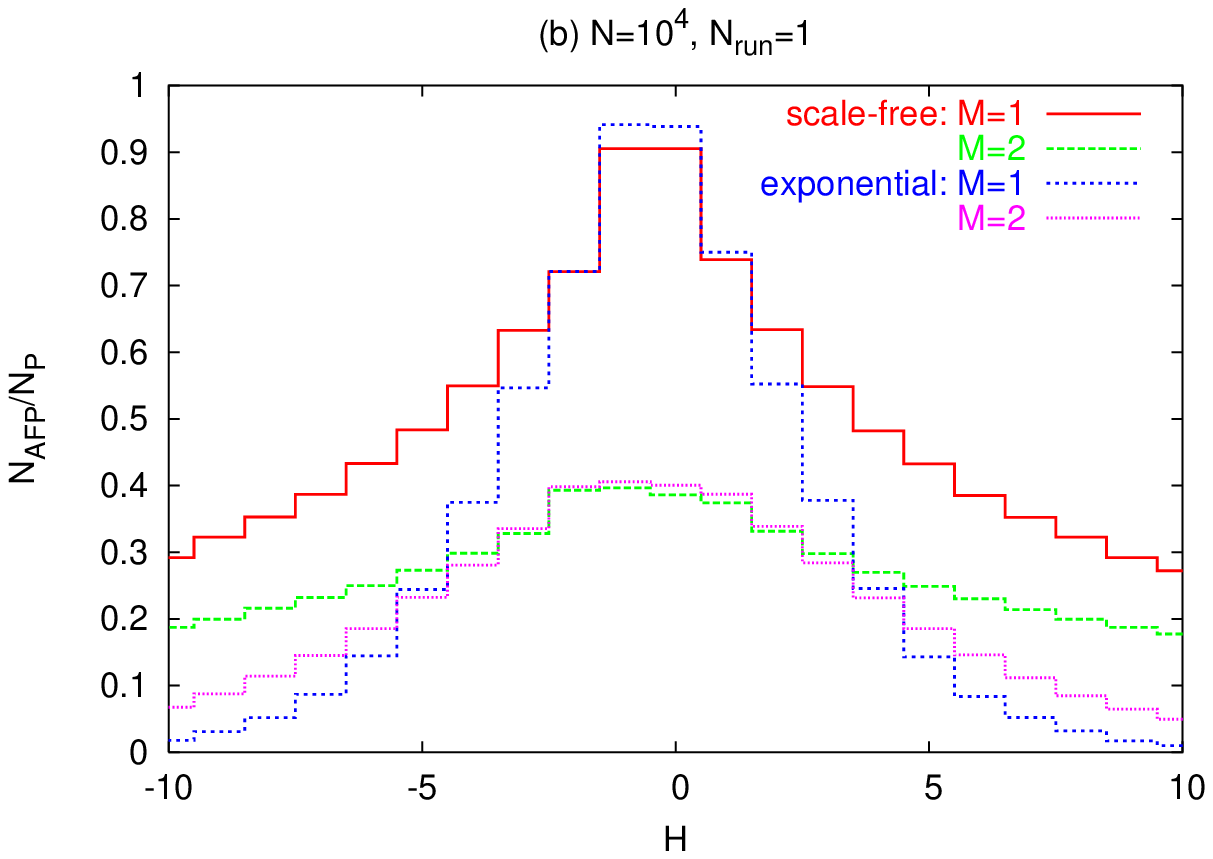}}
\end{center}
\caption{\label{fig-sn-H} The field dependence of
(a) the total number of flipped spins $s(H)/N$ in given field $H$
and (b) density of walls (the number of antiferromagnetic pairs $N_{\text{AFP}}(H)$) normalized to the total number of bonds $N_P$ for various kind of networks.
The field is expressed in $|J|$ units.}
\end{figure}

Results of the dynamics of the remagnetization process are shown in Figs. \ref{fig-mH} and 
\ref{fig-sn-H} for EXN and SFN, $M=1, 2$.
In Fig. \ref{fig-mH}, examples of the hysteresis loop are presented for EXN and SFN, $M=1, 2$. 
The loops are driven in the applied field range from $-H_{\text{max}}-1$ to $H_{\text{max}}+1$,
 where $H_{\text{max}}$ is the saturation field equal to the maximal connectivity 
$k_{\text{max}}$ of the lattice.
All loops show jumps of magnetization at integer field values, as those in Fig. \ref{fig-net-his}.

In Fig. \ref{fig-sn-H}(a) we show the data on the number of flipping spins against the magnetic field, calculated within the same numerical experiment, as in Fig. \ref{fig-mH}. 
In descending field, all the plots in this figure show an asymmetry with respect to the zero field.  Most of spins flip at $H=-1$ for trees, $H=-2$ for the case $M=2$. 
This is a consequence of the connectivity, which is two times smaller for trees in the average.

The results shown in Fig. \ref{fig-sn-H}(b) are equivalent to a plot of the interaction energy against field.
Each antiferromagnetic pair (AFP) gives the contribution $J$ to the energy.
Then the energy of interaction can be written approximately as $J(2N_{\text{AFP}}-MN)$. 
Again we see that higher connectivity enhances the amount of frustration.
We note that the width of the presented curve is much larger for SFN than for EXN.
This is a consequence of the fact that a node with maximal connectivity has 
much more links for the scale-free graphs.

\begin{figure*}
\begin{center}
\resizebox{0.80\hsize}{!}{\includegraphics{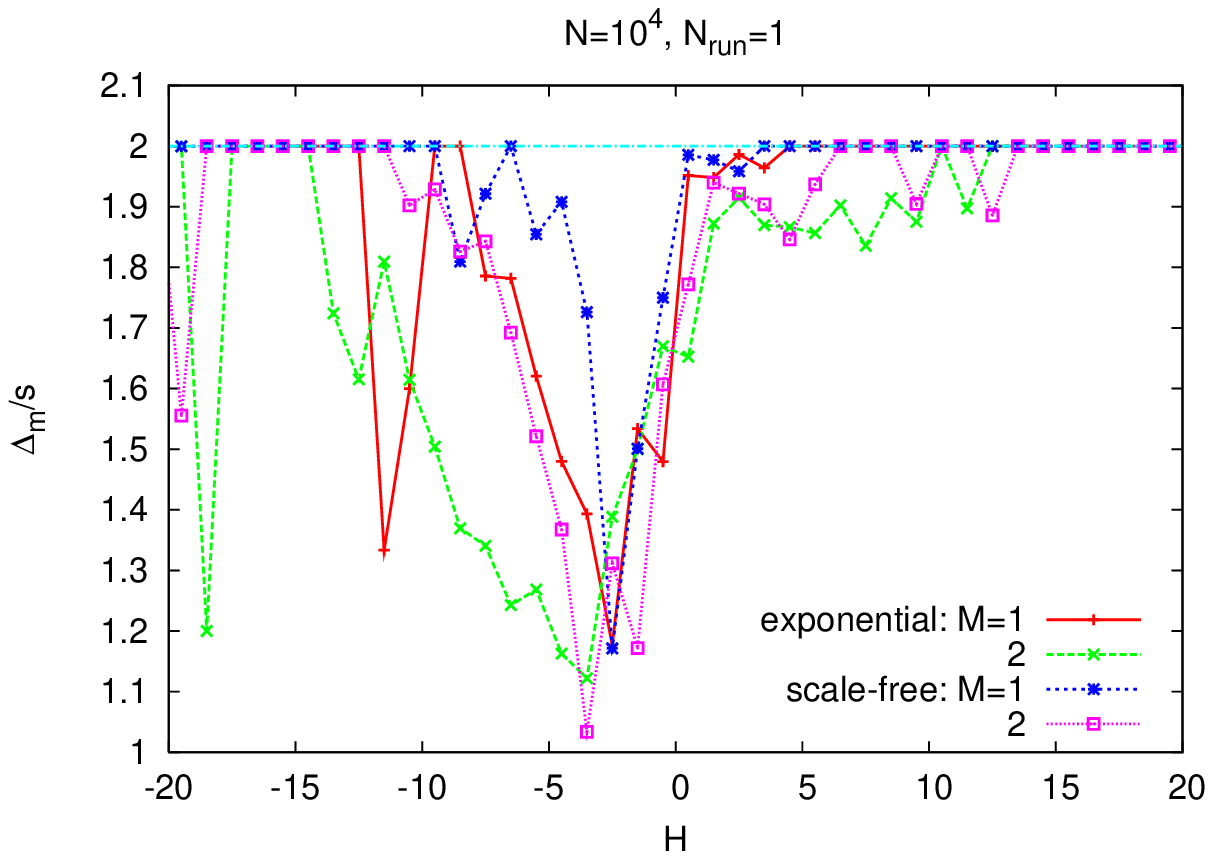}}
\end{center}
\caption{\label{fig-deltam2s} Ratio of the magnetization change $\Delta_m$ and the number $s$ of the flipped spins within an avalanche.}
\end{figure*}

After each field update an  avalanche of spin flips propagates until the energy 
is minimized for the applied  field value.
The distribution of avalanches is integrated along the hysteresis loop, 
and averaged over $N_{\text{run}}$ networks of given kind and size.
This method has been applied both in simulations \cite{vipl} and experiment in
 disordered magnetic systems \cite{spas}. 
The underlying physical mechanism  in these systems is the pinning of domain walls 
by quenched disorder of the network structure. It leads to the Barkhausen noise of 
flipping of magnetic domains \cite{BN}. The role of disorder in the fractal nature 
of the 
observed Barkhausen noise has not been fully understood \cite{disorder_BN}. 
The idea that the underlying processes of   self-\-or\-ga\-ni\-zed criticality \cite{soc} are 
responsible for the occurrence of the scale-free distributions of a\-va\-lan\-ches could not be 
proved in the general case.  Often the observed  finite avalanche cut-offs can be related to
 the finite sizes of the domains. Here we would like to stress that the controversy about 
the domain sizes 
does not appear in our networks. The only relevant scale is the network size. 
Likewise, the concept of  domain walls has a different meaning, as discussed below.
In our numerical experiment we can make  distinction  between the actual change of 
the magnetization $\Delta_m$ and
 the number $s$ of flipped spins within an avalanche $s$, which can be larger 
than $\Delta_m$ (see Fig \ref{fig-deltam2s}). 

\begin{figure*}
\begin{center}
\resizebox{0.45\hsize}{!}{\includegraphics{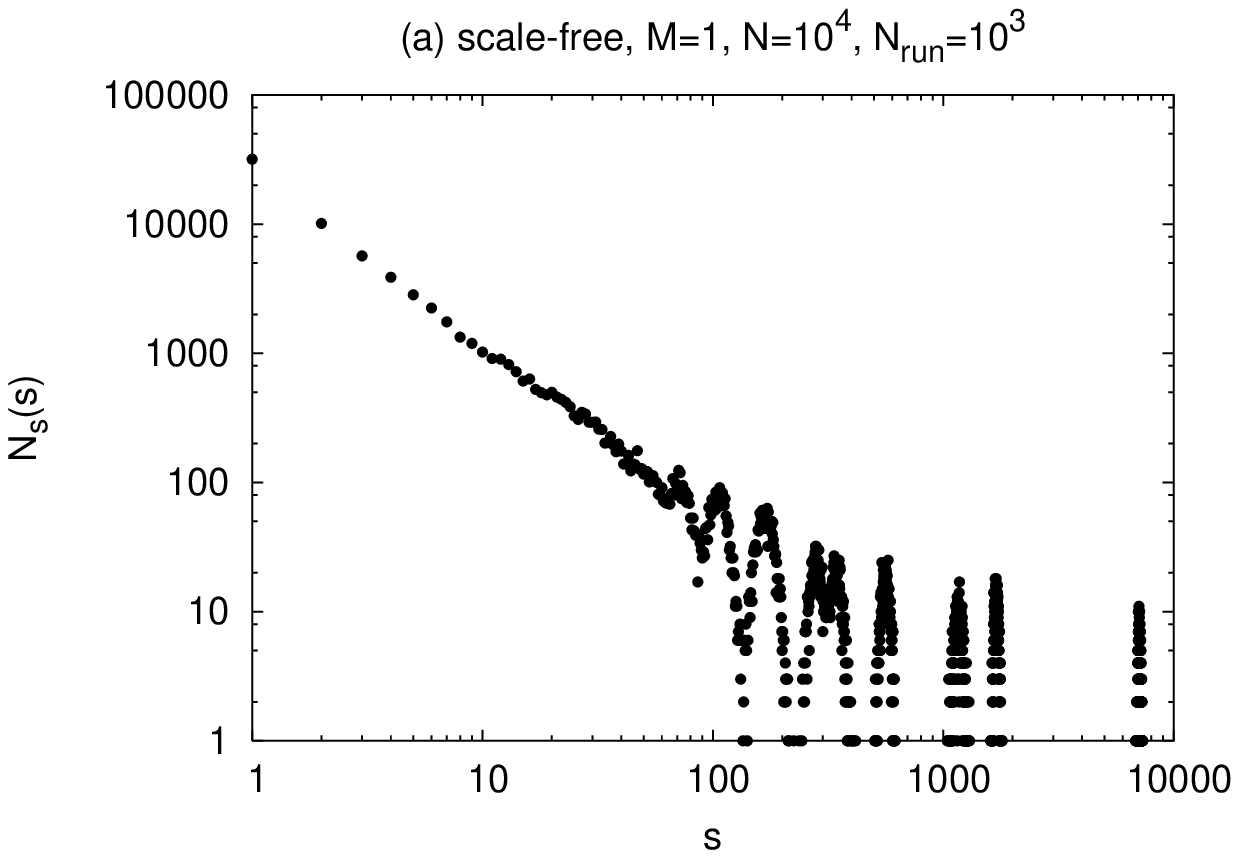}}
\resizebox{0.45\hsize}{!}{\includegraphics{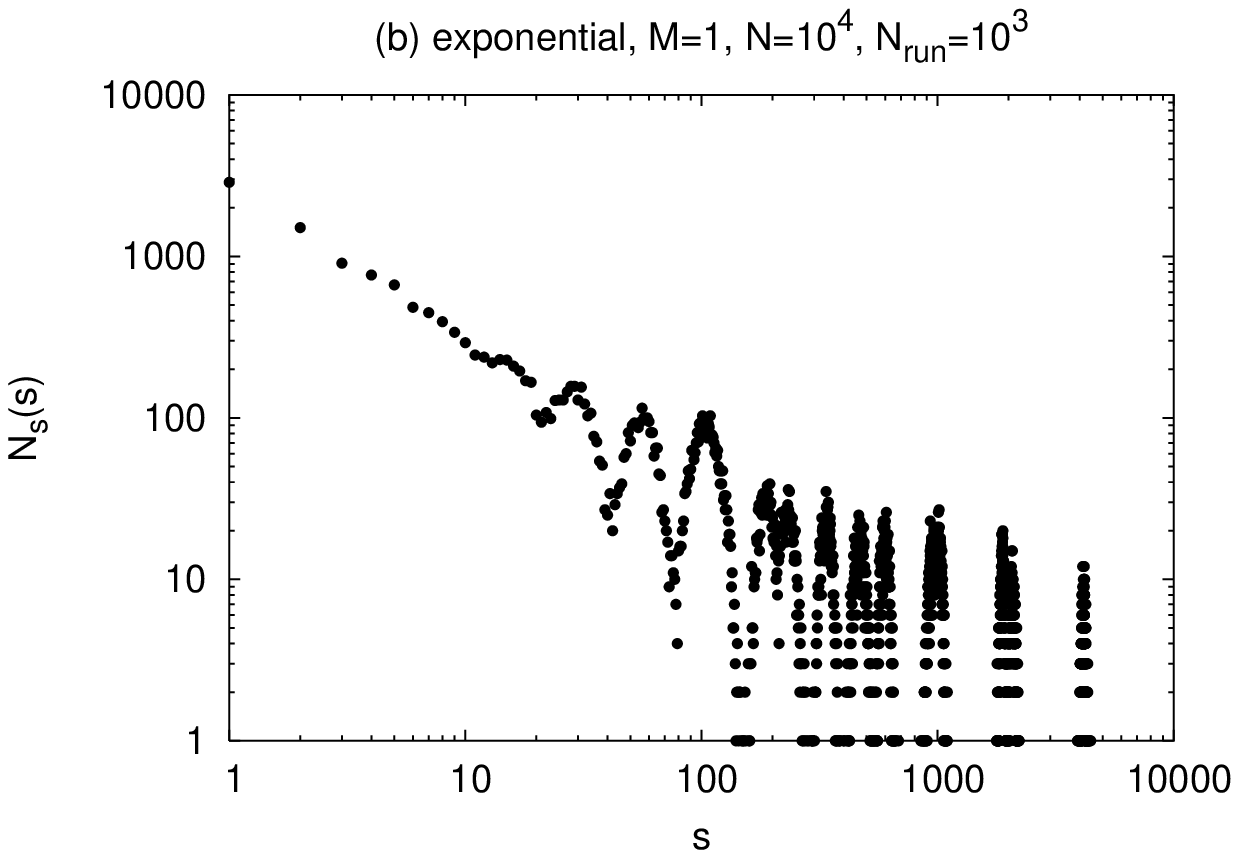}}\\
\resizebox{0.45\hsize}{!}{\includegraphics{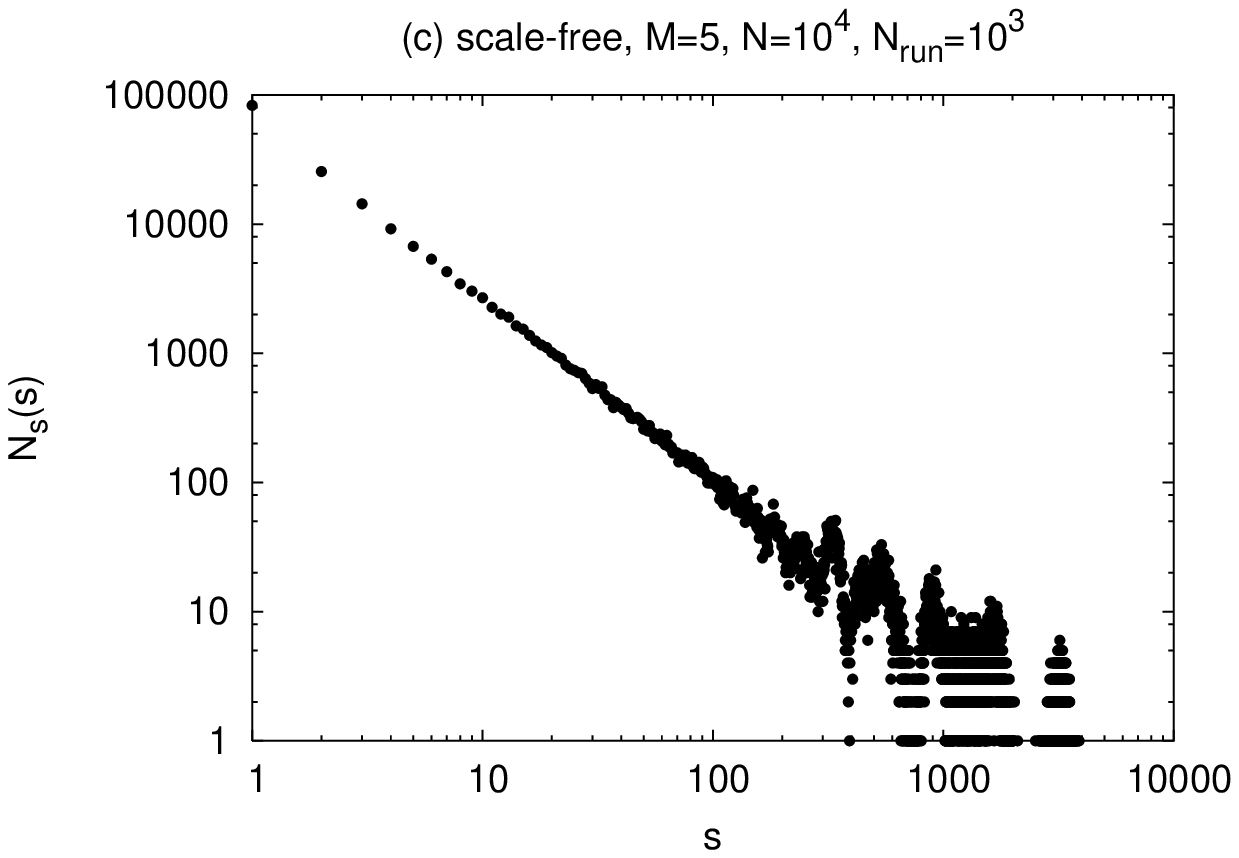}}
\resizebox{0.45\hsize}{!}{\includegraphics{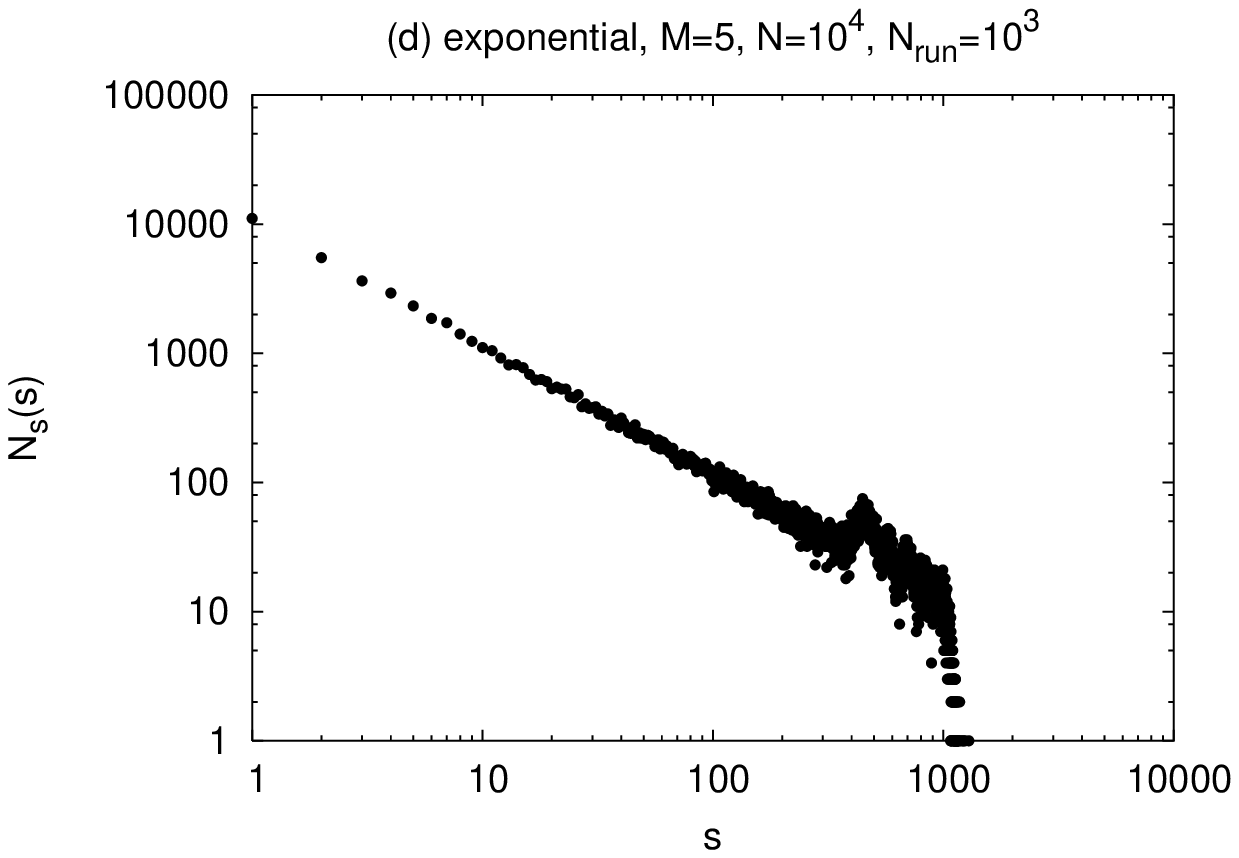}}\\
\resizebox{0.45\hsize}{!}{\includegraphics{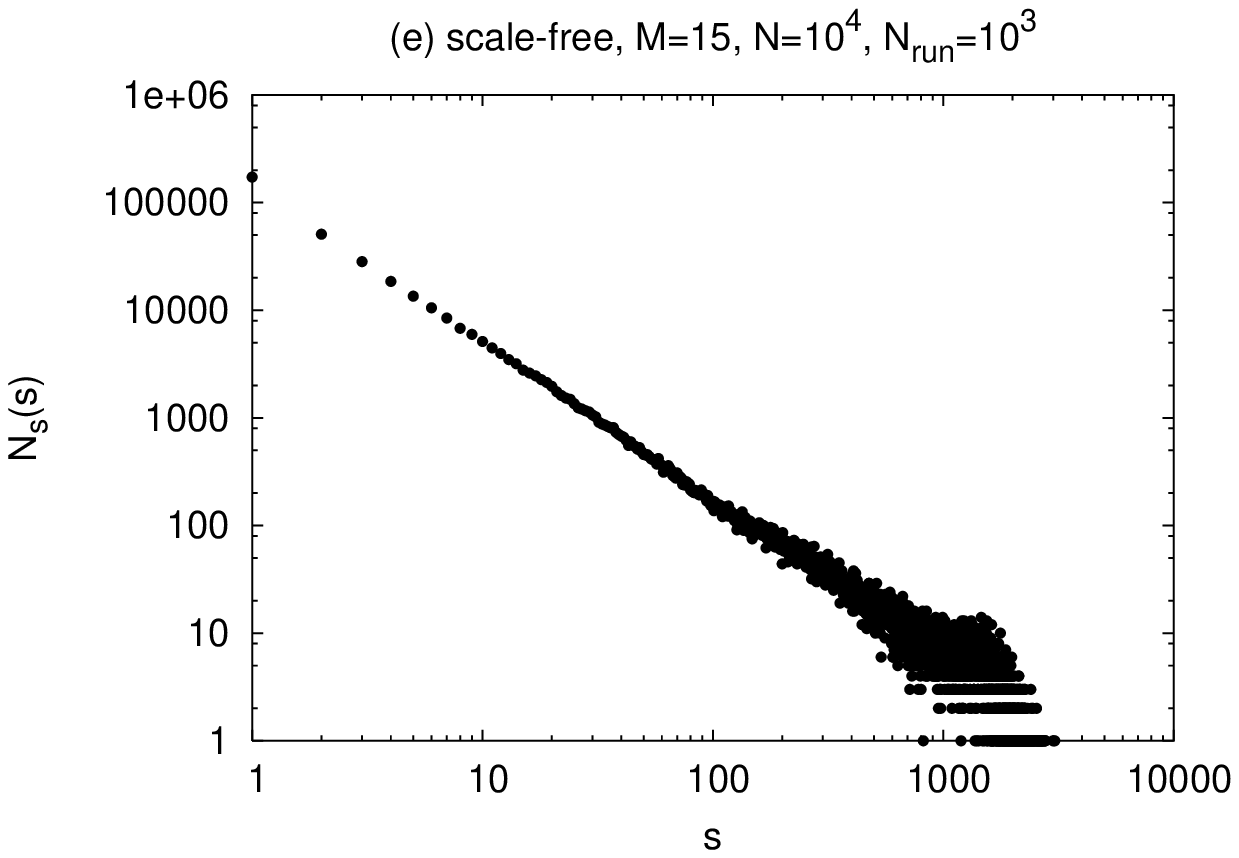}}
\resizebox{0.45\hsize}{!}{\includegraphics{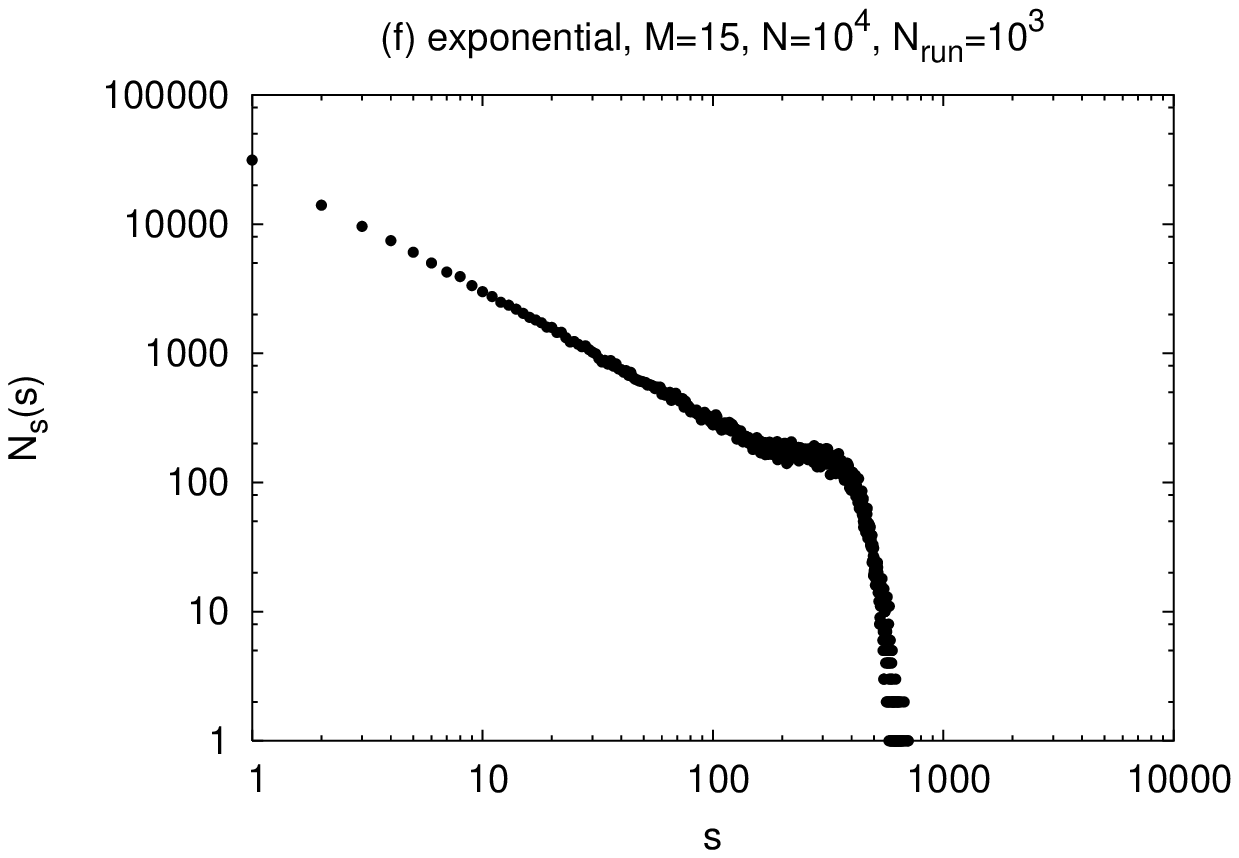}}
\end{center}
\caption{\label{fig-Ns} Histograms of size $s$ of avalanches during hysteresis experiment.}
\end{figure*}

The results for the distribution of avalanches $N_s(s)$ for different kinds of networks are shown in 
Fig. \ref{fig-Ns}. The plots are composed of two parts:
For small values of the avalanche size $s$, the log-log plot is an approximately straight line;
For larger sizes $s$ we observe some oscillations with increasing amplitude.
In the case of trees ($M=1$, absence of frustration) the plots end with a  
well-defined maximum, corresponding to the largest avalanche along the hysteresis loop.
The size of this avalanche is larger in the scale-free tree, compared with the 
exponential tree of the same number of links. For larger values of $M$, the multipeaky character 
of the plots $N_s(s)$ vanishes. However, for small $s$ we observe the power-law behavior
of the spectra $N_s(s)$ in the whole investigated range of $M$, i.e. from 1 to 25.
The values of the effective exponent $\alpha$ --- defined by the relation $N_s(s)\propto s^{-\alpha}$ --- is constant for the exponential networks and it slightly increases with $M$ for the scale-free networks \cite{msc-WA}.

\begin{figure}
\begin{center}
\resizebox{0.80\hsize}{!}{\includegraphics{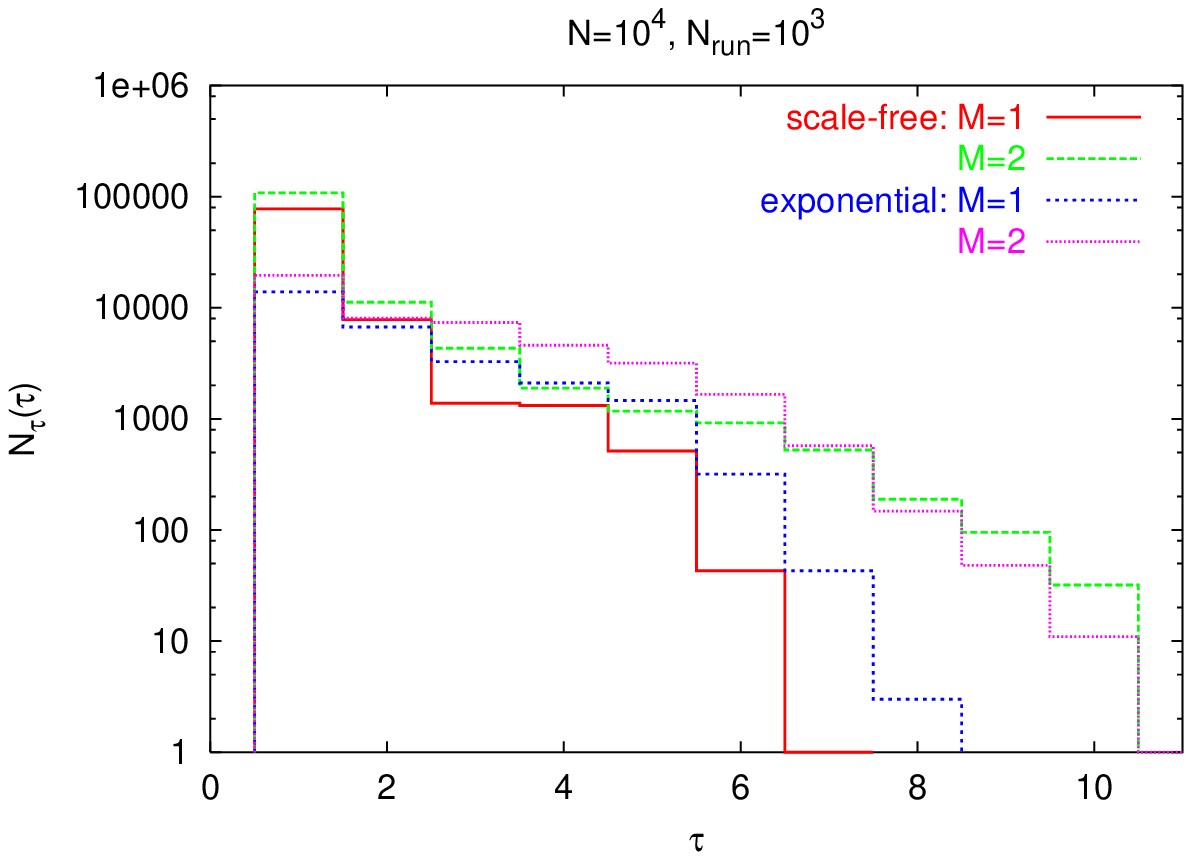}}
\end{center}
\caption{\label{fig-Ntime} Histograms of duration time $\tau$ during hysteresis experiment.}
\end{figure}

The distribution of the time $\tau$ of duration of avalanches is shown in Fig. \ref{fig-Ntime}.
Here, one time step is defined as the time necessary to scan all $N$ spins in the system --- 
spin by spin --- to check if they flip or not.
The time of the avalanche duration $\tau$ is defined as the number of time steps which is 
necessary to get a stable state, i.e., when all flips die out at a given field $H$.
As we see, the range of avalanche durations  $\tau$ covers only one order of magnitude.
The reason is the small-world effect \cite{ab}, i.e., short distance  between the nodes
in random graphs.
As a consequence, the obtained range of $\tau$ increases with the network size $N$ only
as  $\log(N)$.
For the networks investigated here the avalanches never take more than nine or ten 
time steps.
The diameters $\ell$ \cite{task} of the networks considered in this work, 
shown in Tab. \ref{tab}, are comparable to the average avalanche duration. 

\begin{figure}
\begin{center}
\resizebox{0.8\hsize}{!}{\includegraphics{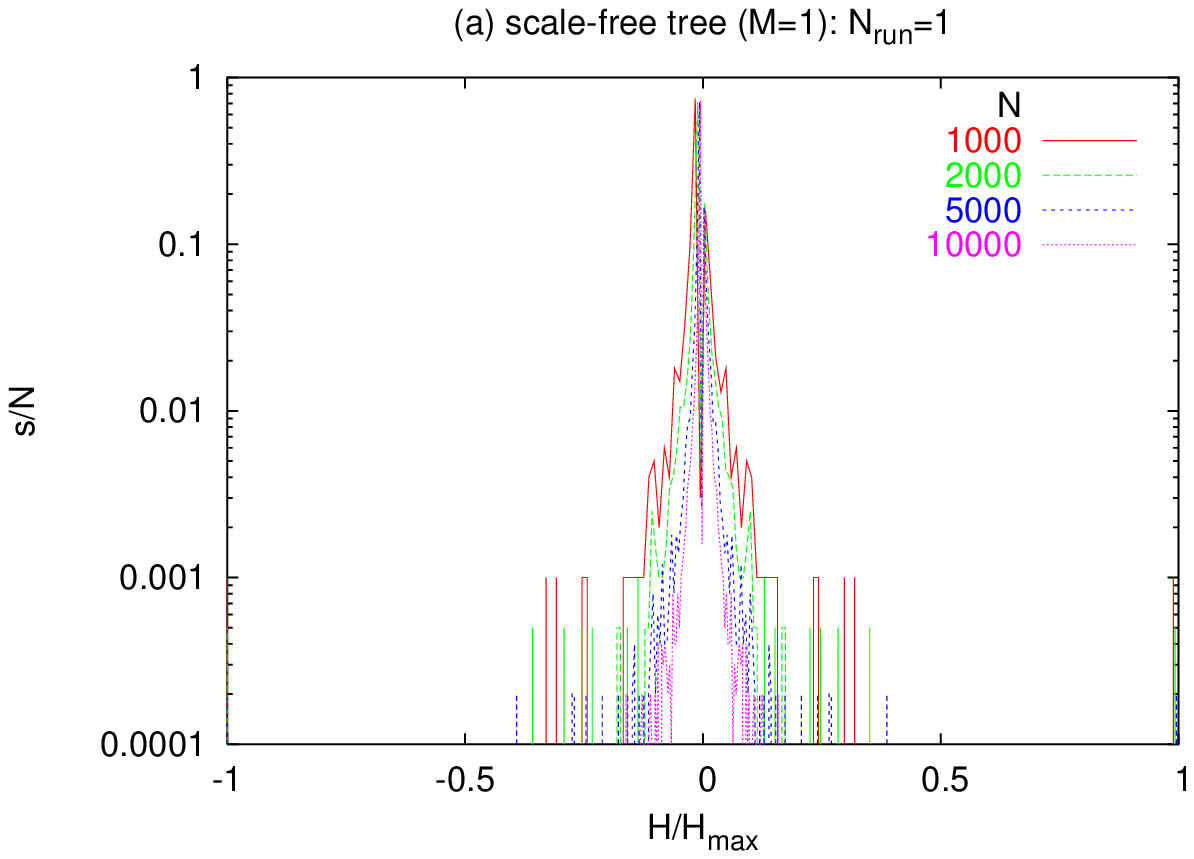}}\\
\resizebox{0.8\hsize}{!}{\includegraphics{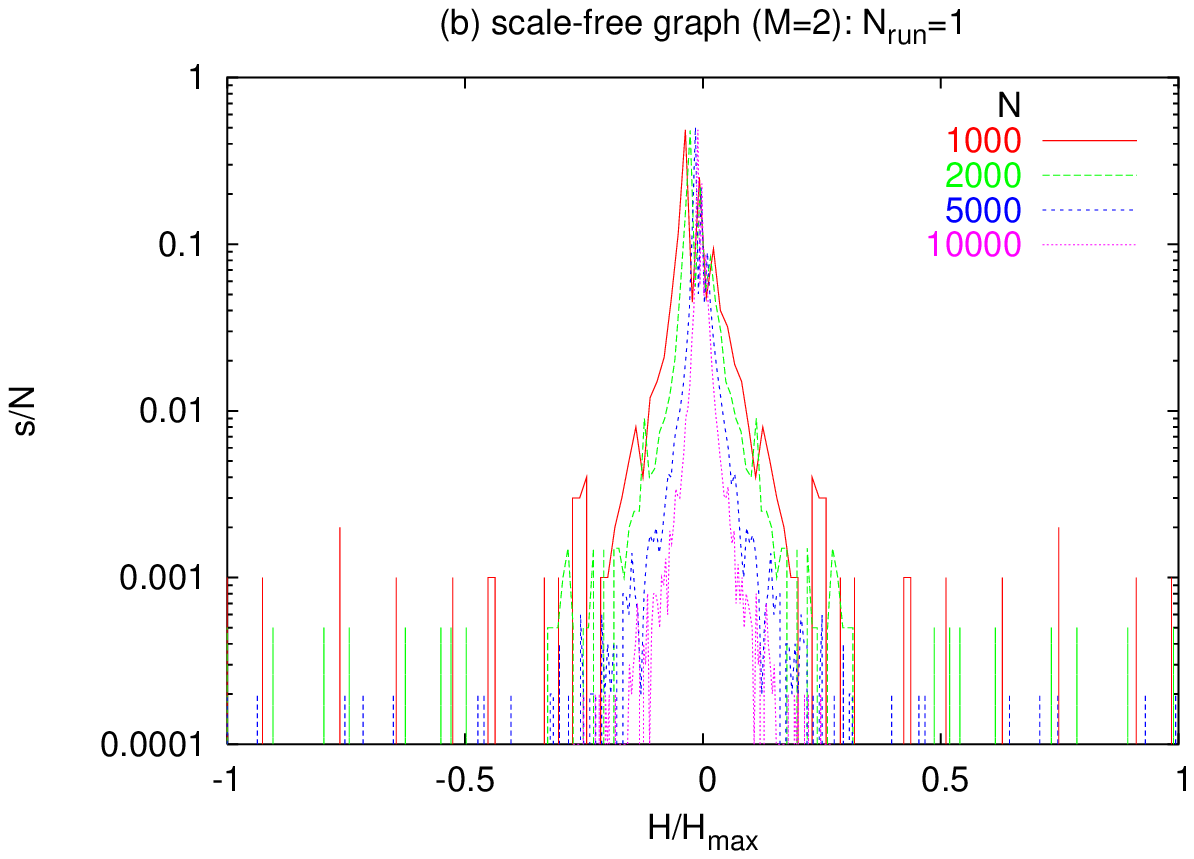}}
\end{center}
\caption{\label{fig-flips} Fraction of flipped spins $s/N$ normalized to the network size $N$ in increasing field $H$ for various values $N$ in the scale-free networks with (a) $M=1$ and (b) $M=2$.}
\end{figure}

The data in Fig. \ref{fig-sn-H}(a) for SFN are compared with those in  
smaller graphs for $M=1$ 
[Fig. \ref{fig-flips}(a)] and $M=2$ [Fig. \ref{fig-flips}(b)].
In these figures the field is normalized to $H_{\text{max}}$, which is different for 
different graph size, i.e., $H_{\text{max}}\approx 100, 100, 200$ and $300$ for 
$N=10^3, 2\cdot 10^3, 5\cdot 10^3$ and $10^4$, respectively.
The logarithmic plots in Fig. \ref{fig-flips} 
reveal that the maximal avalanches normalized by number of nodes $N$  do not decay 
with the system size $N$. This means that the largest jump in the magnetization
on the hysteresis curve persists when the network size increases, suggesting that 
true criticality of the hysteresis loop \cite{DD} occurs in the scale-free networks.
Note that for trees, there is a gap of the avalanche spectrum between $H=H_{\text{max}}$ 
and $H=0$. This gap vanishes for $M=2$.
This is visible also in Fig. \ref{fig-Ns}, where the largest avalanches form 
sharp and separate maxima of the curve $N_s(s)$.

\begin{figure}
\begin{center}
\resizebox{0.80\hsize}{!}{\includegraphics{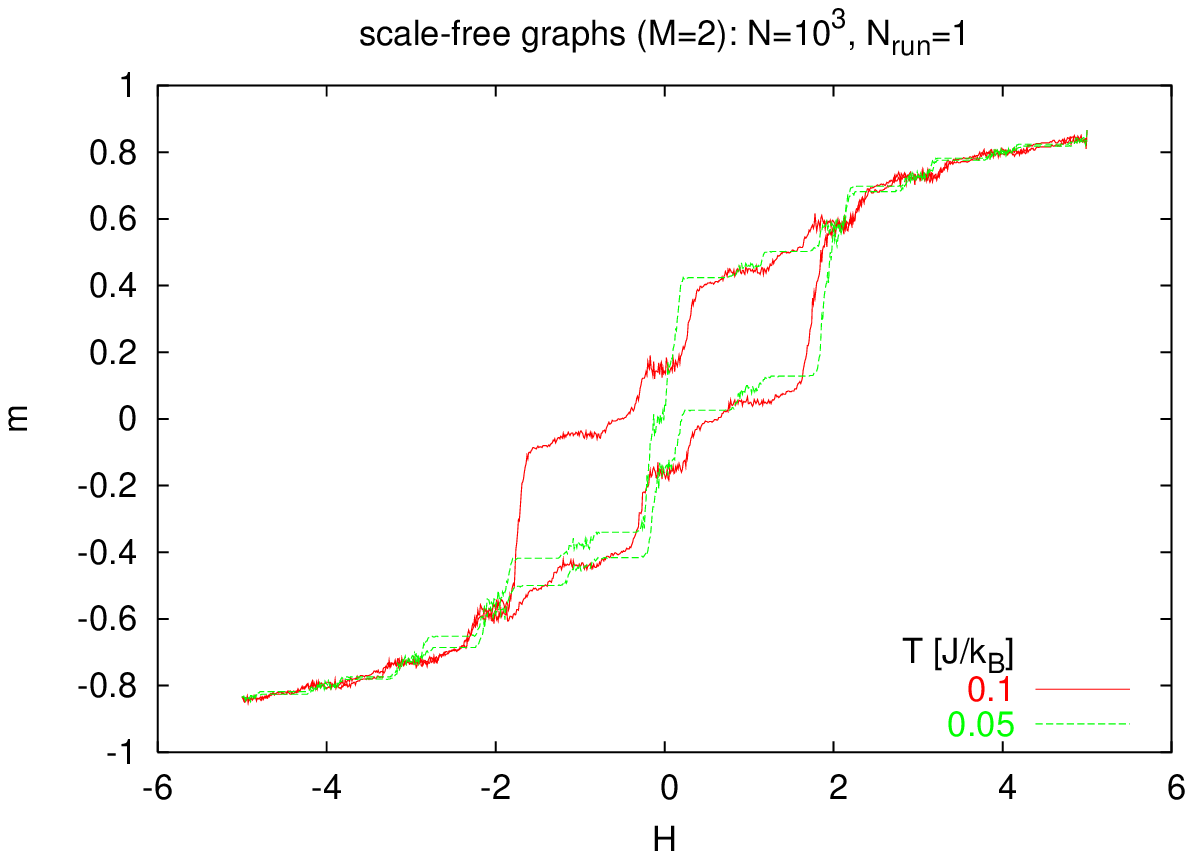}}
\end{center}
\caption{\label{fig-jk} (Color on-line). Hysteresis loop for finite temperature $T=0.05$ and $0.1$ [$|J|/k_B$]. 
The field $H$ is measured in $|J|$.}
\end{figure}

When temperature increases, the hysteresis loops observed in the numerical experiments 
become more noisy, their area decreases and finally vanishes for $T\approx 0.5~[|J|/k_B]$.
Two examples of such loops obtained with Metropolis algorithms \cite{metro} are shown 
in Fig. \ref{fig-jk} for $T=0.05$ and $T=0.1$ [$|J|/k_B$] \cite{msc-JK}.
The avalanche size distribution $N_s(s)$ preserves its power-law character for the above 
values of temperature. 
However, the thermal noise does not allow to separate sharp peaks for large $s$, which 
are visible in the case of $T=0$ (Fig. \ref{fig-Ns}). We note, that the calculations 
for $T>0$ are time-consuming. Still, we can deduce from the results of the simulation  
that the magnetization  curves  of the spin-networks at finite temperature are similar
to those  at $T=0$, 
as long as the thermal energy is much smaller that the energy of the node-node magnetic 
interaction between nearest neighbors.

A typical running for $N=10^4$ and $M=2$ takes about half of minute of CPU time on Intel Itanium 2 (1.5 GHz) processor.
For statistics given here ($N_{\text{run}}=10^3$) it gives over eight hours of CPU time per one network and one branch of hysteresis $m(H)$.

\section{\label{sec5}Discussion and conclusions}

The results show that the connectivity distribution in complex networks with antiferromagnetic
bonds leads to broad distributions of avalanches. In the scale-free networks, the maximal degree
is larger than in the exponential networks of the same size. This difference produces broader
loops for SFN than for EXN. On the other hand, for networks with larger connectivity ($M=2$)
the right parts of histograms of avalanches are irregular, whereas for trees ($M=1$) we observe
fine structure of sharp maxima with zero signal between them. 

It is possible to interpret the size of maximal avalanches $s_{\text{max}}$ in terms of the 
structure 
of the investigated networks. It appears that $s_{\text{max}}$ is comparable with the number
of nodes with minimal degree. On the other hand, $s_{\text{max}}$ happens to be equal to 
the number of spins which flip at field between $H=-M+\delta$ and $H=-M-1+\delta$ for decreasing 
field $H$. Let us trace these rules in detail.

In Fig. \ref{fig-Ns} we see that maximal 
avalanches for SFN trees contain approximately 7100 flipping spins, i.e. 71 percent of the system size.
This value corresponds to the height 0.72 of the maximum in Fig. \ref{fig-sn-H}(a), which happens to 
appear at $H=-|J|$ again for the decreasing field branch. On the other hand, the degree 
distribution for SFN is known 
to be \cite{pk-sfn} 
\begin{equation}
\label{eq-pk-sfn}
P(k\ge M)=\frac{2M(M+1)}{k(k+1)(k+2)}.
\end{equation}
For trees, the minimal degree is $k_{\text{min}}=M=1$, then, for SFN we get 
$P(k=1)=2/3$. These three numbers, in first column of numbers in Table 1 (scale-free, M=1)
are close to each other; our interpretation is that they give approximate information 
on the same process.

The same  expression \eqref{eq-pk-sfn} applies to 
SFN with $M=2$, i.e. $k_{\text{min}}=2$, as the degree cannot be smaller than $M$. 
Calculating  $P(k=2)$ from 
Eq. \eqref{eq-pk-sfn} gives $P(k=2)=1/2$. This is to be compared with the 
size of the maximal avalanche 
for SFN, $M=2$ in Fig. \ref{fig-Ns}, and to the fraction of spins which flip 
at field close to $H=-2 |J|$, as shown in Fig. \ref{fig-sn-H}(a). The numerical values
are given in the second column (scale-free, $M=2$) of Tab. \ref{tab}. As we see, they 
all of them are close to 0.5. Again, this means that the largest avalanche contains spins
at all nodes with $M=2$, and this avalanche happens also at $H=-|J|M$. 

\begin{table}
\caption{\label{tab} Fraction of spins flipped at $H=-M$ [Fig. \ref{fig-sn-H}(a)] and maximal 
avalanche size $s_{\text{max}}/N$ (Fig. \ref{fig-Ns}) as compared with fraction of spins 
with degree equal to $M$ \cite{drm,app,pk-sfn} for various networks.
The effective exponents $\alpha$ and $\beta$ of power law distributions are included for $N_s(s)$ and 
$N_m(\Delta_m)$, respectively.
The last line shows network diameters $\ell$ \cite{task} for $N=10^4$.}
\begin{center}
\begin{tabular}{r cccc} 
\hline \hline
    & \multicolumn{2}{c}{scale-free} & \multicolumn{2}{c}{exponential}\\
$M$                & 1       & 2       & 1       & 2      \\
\hline
$s$ for $H=-M$     & 0.72    & 0.49    & 0.41    & 0.23 \\
$s_{\text{max}}/N$ & 0.73    & 0.54    & 0.44    & 0.25 \\
$P(k=M)$         & 2/3     & 1/2     & 1/2     & 1/3   \\
\hline
$\alpha$           & 1.38    & 1.39    & 0.97    & 1.00 \\
$\beta$            & 1.56    & 1.48    & 0.92    & 0.93 \\
\hline
$\ell$             & 9.1     & 5.1     & 15.6    & 5.1  \\
\hline \hline
\end{tabular}
\end{center}
\end{table}

The same evaluation for EXN, i.e., $0.41$ for $M=1$, and $0.23$ for $M=2$ 
from Fig. \ref{fig-sn-H}(a), when compared with $N_s(s)$
gives a qualitative accordance. As a rule, the size of maximal 
avalanches is comparable with the system size. This results suggests that the 
self-organized criticality (SOC) might be a possible mechanism of the
spin-reversal dynamics in the antiferromagnetic  disorder-free networks. 
We note that for EXN trees, the number of leaves 
of minimal degree $k=1$ can be evaluated from $P(k\ge M)=2^{-k}$ \cite{drm} as $1/2$ 
for $M=1$. 
For $M=2$, the distribution $P(k\ge M)=3/4\cdot(3/2)^{-k}$ \cite{app} 
gives $P(k=2)=1/3$. Comparison of the results for EXN is included in 
third and fourth column of Tab. \ref{tab}. These numbers prove, that in the 
hysteresis experiment, maximal avalanches for the descending field appear when 
$H=-k_{\text{min}}|J|=-M|J|$.
The spins flipping in these avalanches occupy the nodes with minimal degree $M$. 
The observed quantitative differences might be attributed to the finite network
sizes in the numerical calculations; this produces deviations of simulation results from
theoretical formula, which are exact in the thermodynamic limit.

\begin{figure}
\begin{center}
\resizebox{0.45\hsize}{!}{\includegraphics{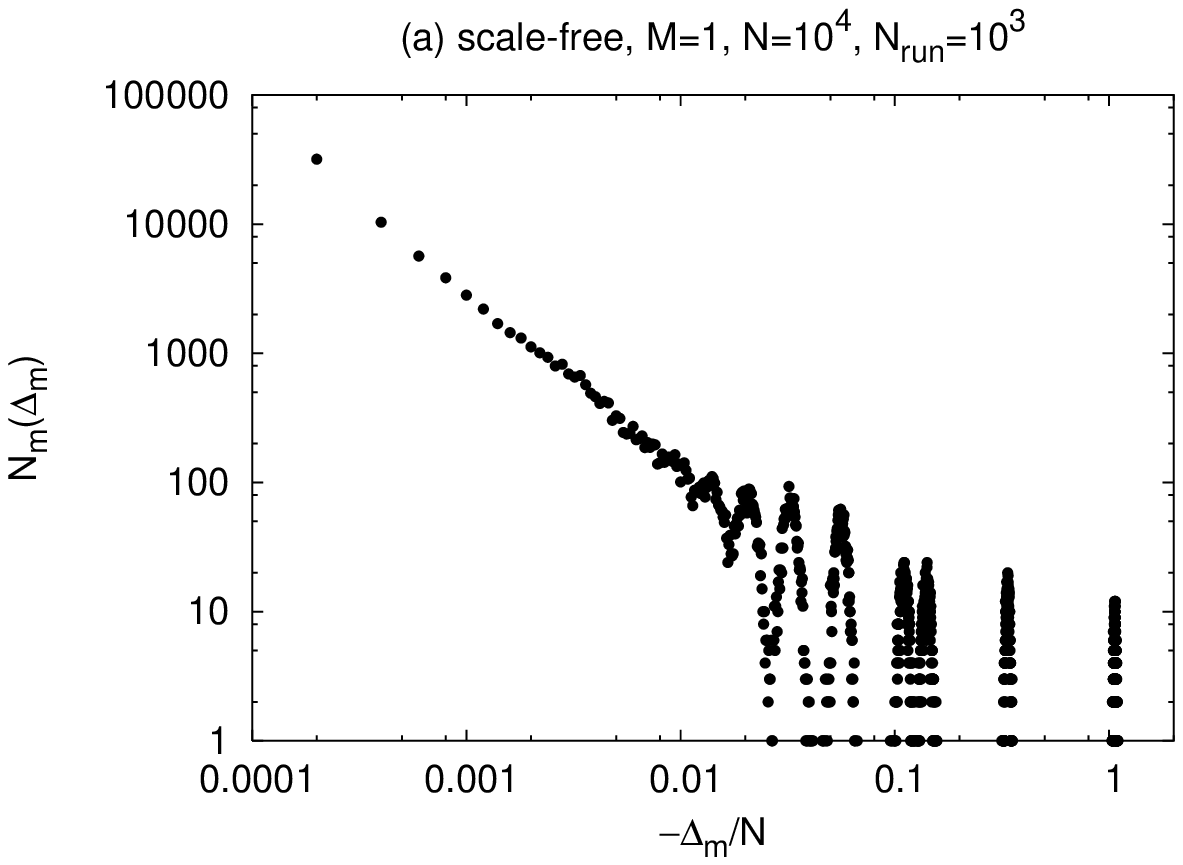}}
\resizebox{0.45\hsize}{!}{\includegraphics{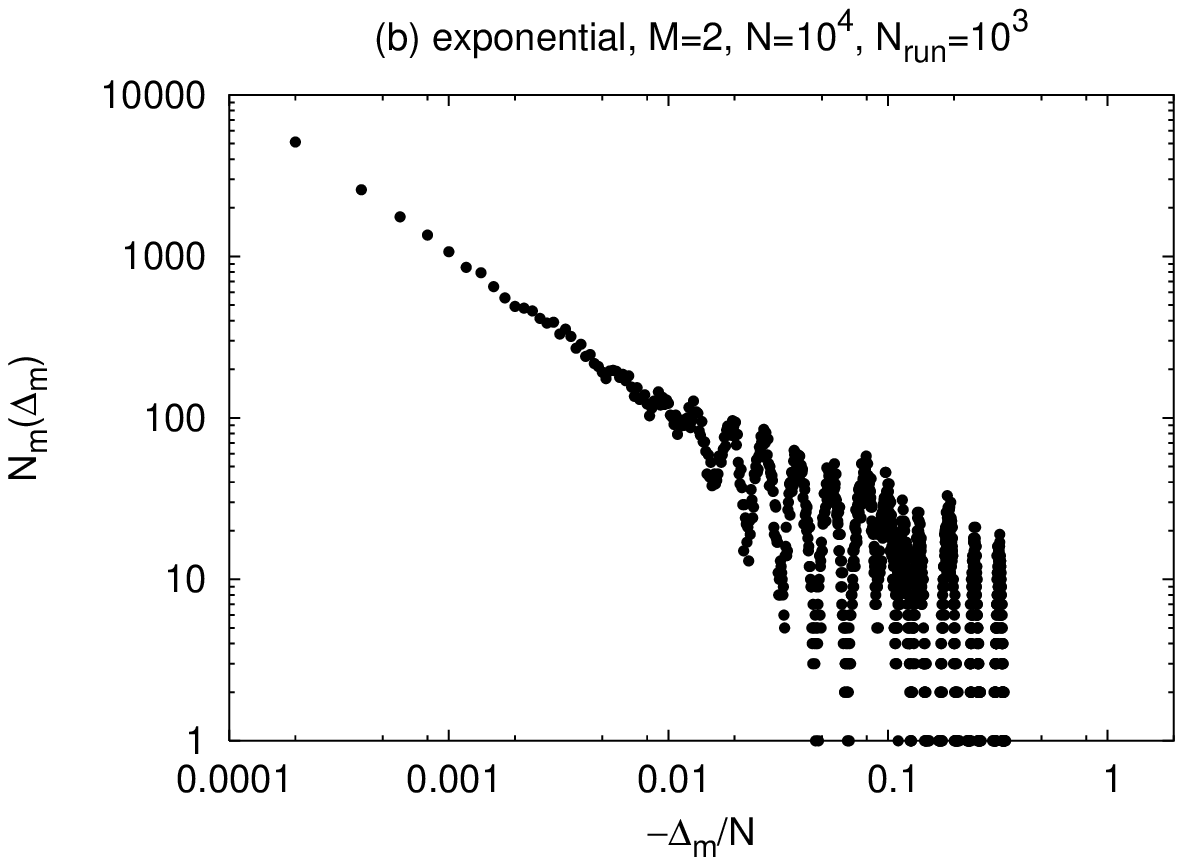}}
\end{center}
\caption{\label{fig-Ndeltam} Histograms of changes of magnetization $\Delta_m$ of avalanches during hysteresis experiment.}
\end{figure}

Left parts of the histograms, i.e.
the spectra for small avalanche sizes, are straight lines in the log-log scale.
This means that $N_s(s)\propto s^{-\alpha}$, with $\alpha$ close to $1.00$ for the 
exponential networks and $\alpha$ slightly increasing with $M$ for the scale-free networks. 
It is not clear if this variation of $\alpha$ for the scale-free networks is a consequence of a 
$M$-dependent finite size effect, or it means just a lack of universality.
We cannot verify it, because for large
networks, the numerical algorithm becomes very time-consuming. 
However, this scaling behavior is valid only  
in the range of the avalanche size $s$ not much larger than one order of magnitude 
\cite{BT-etal}.
The latter apply also to the distribution $N(\Delta_m)\propto\Delta_m^{-\beta}$ 
(see Fig. \ref{fig-Ndeltam}).
The effective exponents $\alpha$ and $\beta$ are also shown in Tab. \ref{tab}.
Spins which flip in this way are neither leaves (for $M=1$), nor ``leaves with two 
stems'' (for $M=2$), but they form ``central'' (here the oldest) parts of the networks.
The results suggest that the spin flips  in the better connected parts of the network
form small avalanches with power-law distributions.

Basing on these results we expect that avalanches and critical behavior may occur
in  various  systems 
with broad connectivity distributions where some negative feedback exists between 
local degrees of freedom. In our spin-networks it is the antiferromagnetic character 
of interaction which causes this 
kind of feedback. More direct interpretation depends on particular systems. 
In particular, avalanches
seem possible in complex magnetic systems as quantum magnetic cellular automata 
with magnetostatic
interaction between particular monodomain elements \cite{cow}. The size of avalanches 
is limited only by the size of the network. Then, this kind of collective flipping 
is related to an instability of the system. Design of an appropriate spatial structure,
in particular by increasing clustering of the networks \cite{BT-etal}  
allows to tune and control the magnetization curve.

Concluding, our numerical results indicate that the positions of the largest peaks in the 
spectra of avalanches, 
which measure the size of the largest avalanches, agree well with the numbers of spins 
flipped at the field equal to the connectivity parameter $M$.
The latter can be interpreted as the numbers of nodes with minimal degree for networks with 
different topologies and connectivities.
These results agree qualitatively with known theoretical formula for the degree distributions.

\ack
K.M. is grateful for hospitality at JSI, Ljubljana.
Part of calculations was carried out in ACK\---CY\-FRO\-NET\---AGH.
The machine time on HP Integrity Superdome is financed by the Polish Ministry of Education 
and Science under grant No. MNiI/HP\_I\_SD/AGH/047/2004.



\end{document}